%% file: main.tex
\documentclass[twocolumn]{aastex63}
\pdfoutput=1 %for arXiv submission
\usepackage{amsmath,amstext}
\usepackage[T1]{fontenc}
\usepackage{apjfonts} 
\usepackage[figure,figure*]{hypcap}
\usepackage{longtable}
\usepackage{threeparttable}

\usepackage{amssymb}
\usepackage{amsmath}
\usepackage{fontawesome}
\usepackage{gensymb}
\usepackage{mathrsfs}
\usepackage{upgreek}
\usepackage{fontenc}
\usepackage{hyperref}
\usepackage{float}
\usepackage{tabularx}

\newcommand{\ergs}{\text{erg s}\ensuremath{^{-1}}}

\newcommand{\kms}{\text{km s}\ensuremath{^{-1}}}

\renewcommand{\d}{\ensuremath{\rm d}}

\shorttitle{CLASS: Coronal Line Activity \rev{Spectroscopic Survey}}
\shortauthors{M. Reefe et al.}

\graphicspath{{./}{figures/}}

\newcommand{\rev}{}

\begin{document}

% \title{A Large Scale Survey of Galaxies with Coronal Line Emission Selected from the Sloan Digital Sky Survey}
% tentative title; Michael, maybe you can think of a flashier one.
% \title{A CLASS Project: Searching for coronal line needles in the SDSS Haystack}
\title{CLASS: Coronal Line Activity \rev{Spectroscopic Survey}}

\author[0000-0003-4701-8497]{Michael Reefe}
\altaffiliation{National Science Foundation, Graduate Research Fellow}
\affiliation{George Mason University, Department of Physics and Astronomy, MS3F3, 4400 University Drive, Fairfax, VA 22030, USA}
\author[0000-0003-2277-2354]{Shobita Satyapal}
\affiliation{George Mason University, Department of Physics and Astronomy, MS3F3, 4400 University Drive, Fairfax, VA 22030, USA}
\author[0000-0003-3432-2094]{Remington O. Sexton}
\affiliation{George Mason University, Department of Physics and Astronomy, MS3F3, 4400 University Drive, Fairfax, VA 22030, USA}
\affiliation{U.S. Naval Observatory, 3450 Massachusetts Avenue NW, Washington, DC 20392-5420, USA}
\author[0000-0003-3152-4328]{Sara M. Doan}
\affiliation{George Mason University, Department of Physics and Astronomy, MS3F3, 4400 University Drive, Fairfax, VA 22030, USA}
\author[0000-0002-4902-8077]{Nathan J. Secrest}
\affiliation{U.S. Naval Observatory, 3450 Massachusetts Avenue NW, Washington, DC 20392-5420, USA}
\author[0000-0003-1051-6564]{Jenna M. Cann}
\altaffiliation{NASA Postdoctoral Program}
\affiliation{NASA Goddard Space Flight Center, 8800 Greenbelt Rd, Greenbelt, Maryland 20771 USA}

\correspondingauthor{Michael Reefe}
\email{mreefe@gmu.edu}

\begin{abstract}

We conduct the first systematic survey of a comprehensive set of the twenty optical coronal lines in the spectra of nearly 1 million galaxies observed by the Sloan Digital Sky Survey (SDSS) Data Release 8 catalog.  This includes often overlooked high ionization-potential (IP) lines such as [\ion{Ar}{10}] $\lambda$5533, [\ion{S}{12}] $\lambda$7609, [\ion{Fe}{11}] $\lambda$7892, and [\ion{Fe}{14}] $\lambda$5303.  We find that, given the limited sensitivity of SDSS, strong coronal line emission is extremely rare, with only $\sim 0.03$\% of the sample showing at least one coronal line, significantly lower than the fraction of galaxies showing optical narrow line ratios  ($\sim 17$\%) or mid-infrared colors ($\sim 2$\%) indicative of nuclear activity. The coronal line luminosities exhibit a large dynamic range, with values ranging from $\sim10^{34}$ to $10^{42}$ \ergs.   We find that a vast majority ($\sim 80$\%) of coronal line emitters in dwarf galaxies ($M_* < {9.6} \times 10^9$~M$_\odot$) do not display optical narrow line ratios indicative of nuclear activity, in contrast to higher mass galaxies ($\sim 17$\%). Moreover, we find that the highest ionization potential lines are preferentially found in lower mass galaxies.  These findings are consistent with the theory that lower mass black holes found in lower mass galaxies produce a hotter accretion disk, which in turn enhances the higher ionization coronal line spectrum.  Future coronal line searches with 30~m class telescopes or JWST may provide a pathway into uncovering the intermediate mass black hole population.
\end{abstract}


\section{Introduction}

High ionization fine-structure lines have been known to exist in the optical spectra of nearby active galactic nuclei (AGNs) for almost sixty years \citep[e.g.,][]{1968ApJ...151..807O,1970ApJ...161..811N,1978ApJ...221..501G,1984MNRAS.211P..33P, 1988AJ.....95...45A, 2002MNRAS.329..309P, 2017MNRAS.467..540L,  2018ApJ...858...48M}. These so-called ``coronal lines'', first observed in the solar corona, arise from collisionally excited forbidden fine-structure transitions in highly ionized species, with ionization potentials that extend well beyond the Lyman limit to several hundred electron volts.  This corresponds to the so-called ``Big Blue Bump'' thought to arise from the accretion disk emission \citep[e.g.,][]{1969Natur.223..690L,1984ApJ...278..558M,1985MNRAS.216...63N,1992AJ....103.1084P} and the ``soft excess'' characteristic of AGN soft X-ray spectra \citep{1999agnc.book.....K, 2016A&A...588A..70B}. While systematic surveys of the strong emission lines seen in the optical spectra of galaxies have now been carried out on millions of galaxies and their diagnostic potential in studying star formation and accretion activity well-established \citep[e.g.,][]{2003MNRAS.346.1055K,2004MNRAS.351.1151B}, no such large scale systematic survey of the much weaker coronal lines has been carried out on similarly large samples of galaxies. Because of their high ionization potentials and high critical densities ($\sim 10^{7}$--$10^{10}$ cm$^{-3}$), coronal lines are powerful probes of the highly ionized and dense gas in galaxy centers. They have the potential to identify elusive AGNs missed by optical narrow line ratios and mid-infrared color selection which suffer from contamination from star formation in the host galaxy \citep[e.g.,][]{2009MNRAS.398.1165G,2015ApJ...811...26T, 2018ApJ...858...38S, 2021ApJ...906...35S}, and may actually be a powerful tool in identifying accreting intermediate mass black holes \citep[e.g.,][]{2018ApJ...861..142C, 2021ApJ...912L...2C, 2021ApJ...910....5M, 2021ApJ...922..155M}.

Coronal lines can indirectly probe the spectral energy distribution (SED) of the ionizing radiation field from the Lyman limit up to several hundred electron volts, a region of the electromagnetic spectrum that is observationally inaccessible because of Galactic and intrinsic absorption \citep[e.g.,][]{1996A&A...315L.109M,2000ApJ...536..710A}. As a result, they can potentially be used to constrain accretion disk models, and even possibly the mass of the black hole \citep{2018ApJ...861..142C, 2022MNRAS.510.1010P}. The resulting intrinsic SED is also of interest in studying the effect of the AGN on its environment, and the line ratios can also be used to infer the gas density and temperature of the highly ionized and dense gas close to the central supermassive black hole (SMBH). Finally, because the line profiles often show asymmetrical profiles indicative of outflows, with outflow properties often showing correlations between ionization potential and critical density \citep[e.g.,][]{1997A&A...323..707E, 2006ApJ...653.1098R,2021ApJ...911...70B}, coronal lines can be used to study outflows in the most highly ionized gas near their launch origin, a crucial input in understanding AGN feedback and its impact on star formation in the host galaxy.

With the advent of the next generation of 30~m class extremely large telescope (ELTs), unprecedented sensitivity and spatial resolution will become possible within the coming decade. Large scale surveys of the fainter coronal lines will be feasible, resulting in transformative advances in our understanding of the highly ionized gas in galaxy centers with extremely high spatial resolution. 
% The optical regime in particular is where most large-scale surveys are focused, and it is where we can achieve the highest spatial resolution and greatest sensitivities from the ground. 
While most past optical and near-infrared coronal line studies have exclusively targeted AGNs \citep{2002ApJ...579..214R,2006A&A...457...61R,2011ApJ...743..100R,2011ApJ...739...69M,2017MNRAS.467..540L,2018ApJ...858...48M}, there have been a few studies aimed at the general galaxy population. In a large scale search for [\ion{Fe}{10}] $\lambda$6374 emission,  \citet{10.1111/j.1365-2966.2009.14961.x} identified 63 galaxies with detections, all of which were previously identified as AGNs. Recently, \citet{2021ApJ...920...62N} conducted an extensive search for coronal line emission in integral field data and found 10 coronal line emitters, 7 of which were known AGNs, all showing spatially extended emission. In other recent work, \citet{2021ApJ...922..155M} searched for [\ion{Fe}{10}] $\lambda$6374 emission in a large sample of dwarf galaxies, finding convincing evidence for previously hidden AGNs in the dwarf galaxy population. These studies have hinted at the diagnostic potential of coronal lines in the general galaxy population and have provided tantalizing evidence for their power in uncovering previously unidentified AGNs, but no study thus far has explored all of the optically accessible coronal lines in any sample, and studies of the comprehensive set of optical coronal lines in large samples of galaxies does not yet exist. The full diagnostic potential of coronal lines from current optical surveys has therefore not yet been extracted.  

In this work, we introduce the Coronal Line Activity \rev{Spectroscopic Survey} (CLASS) \rev{for the Sloan Digital Sky Survey (SDSS)} catalog. We quantify for the first time the detection statistics for all optical coronal lines \rev{detected in the SDSS spectra of a large sample of galaxies}, and analyze the host galaxy demographics of coronal line emitters compared to those from other galaxy classes, and the sample as a whole. The full CLASS catalog, with an in-depth description of our methodology and all coronal line fluxes and kinematics, is presented in Reefe et al. 2022b. Throughout this paper, we assume a flat $\Lambda$CDM cosmology with $H_0=70$~km~s$^{-1}$~Mpc$^{-1}$, $\Omega_m = 0.3$, and $\Omega_\Lambda = 0.7$.

\section{Sample Selection and Methodology}
% text on what produces coronal line emission

\begin{figure*}
    \centering
    \includegraphics[width=\columnwidth]{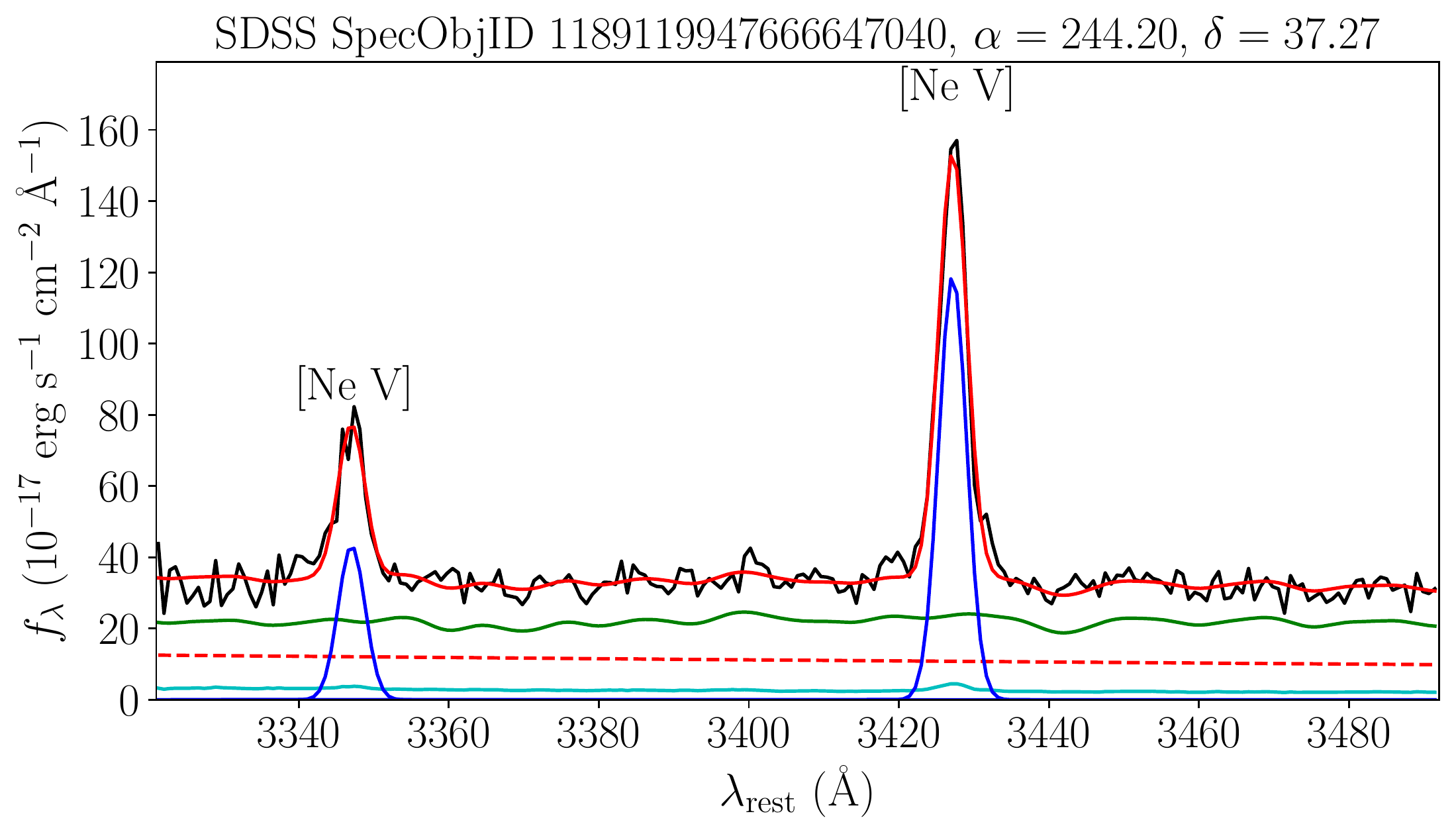}
    \includegraphics[width=\columnwidth]{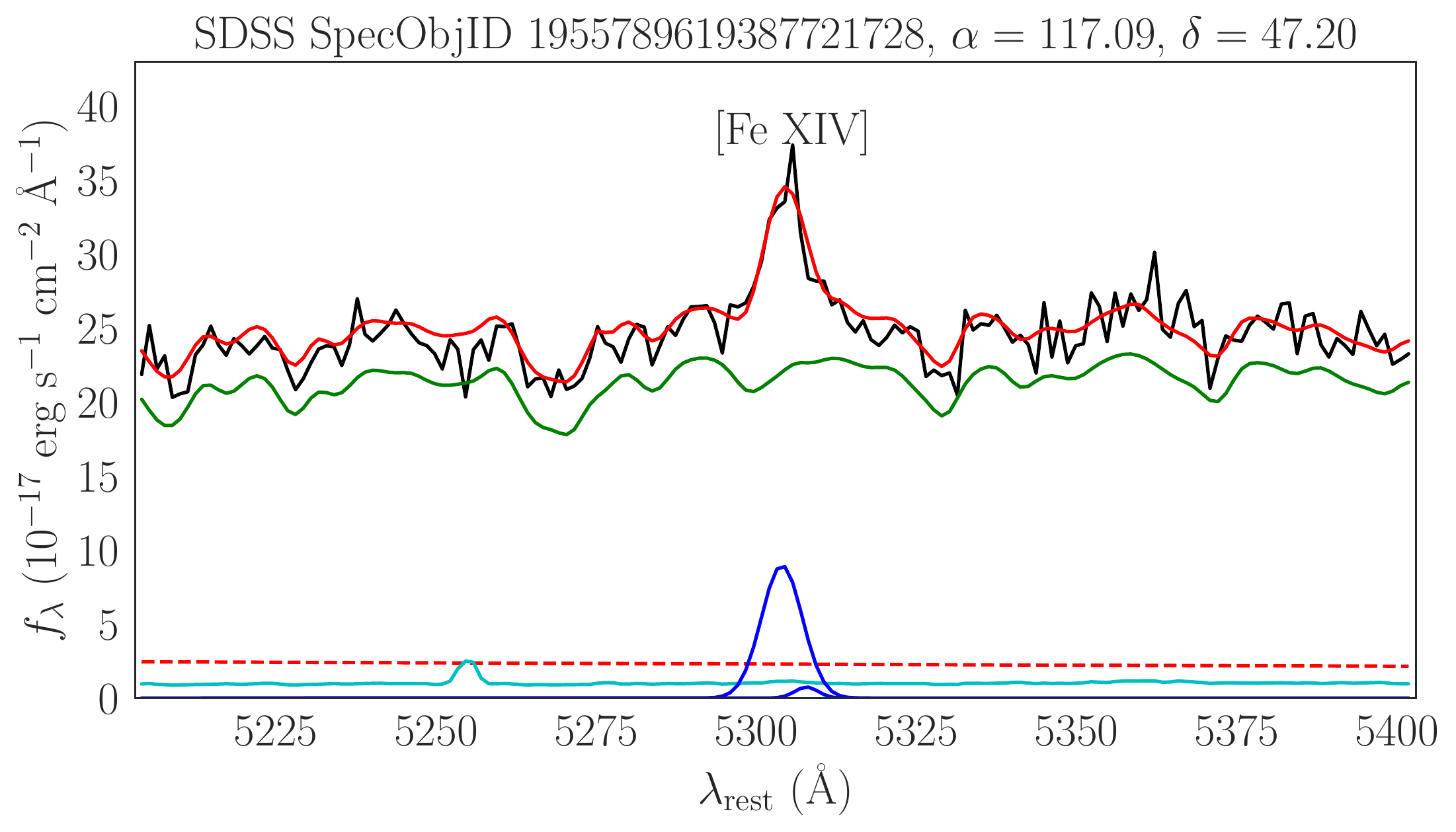}
    \includegraphics[width=\columnwidth]{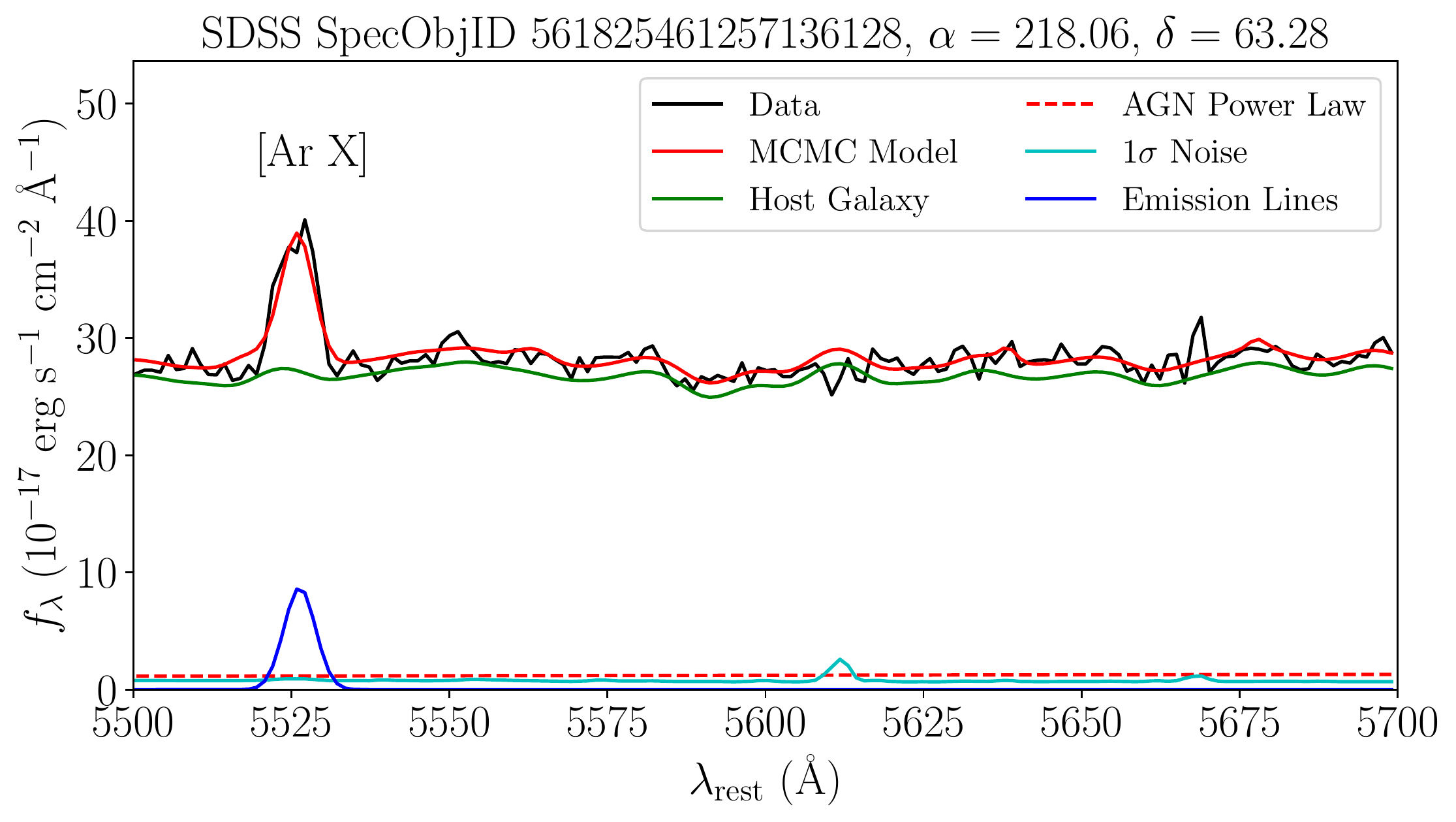}
    \includegraphics[width=\columnwidth]{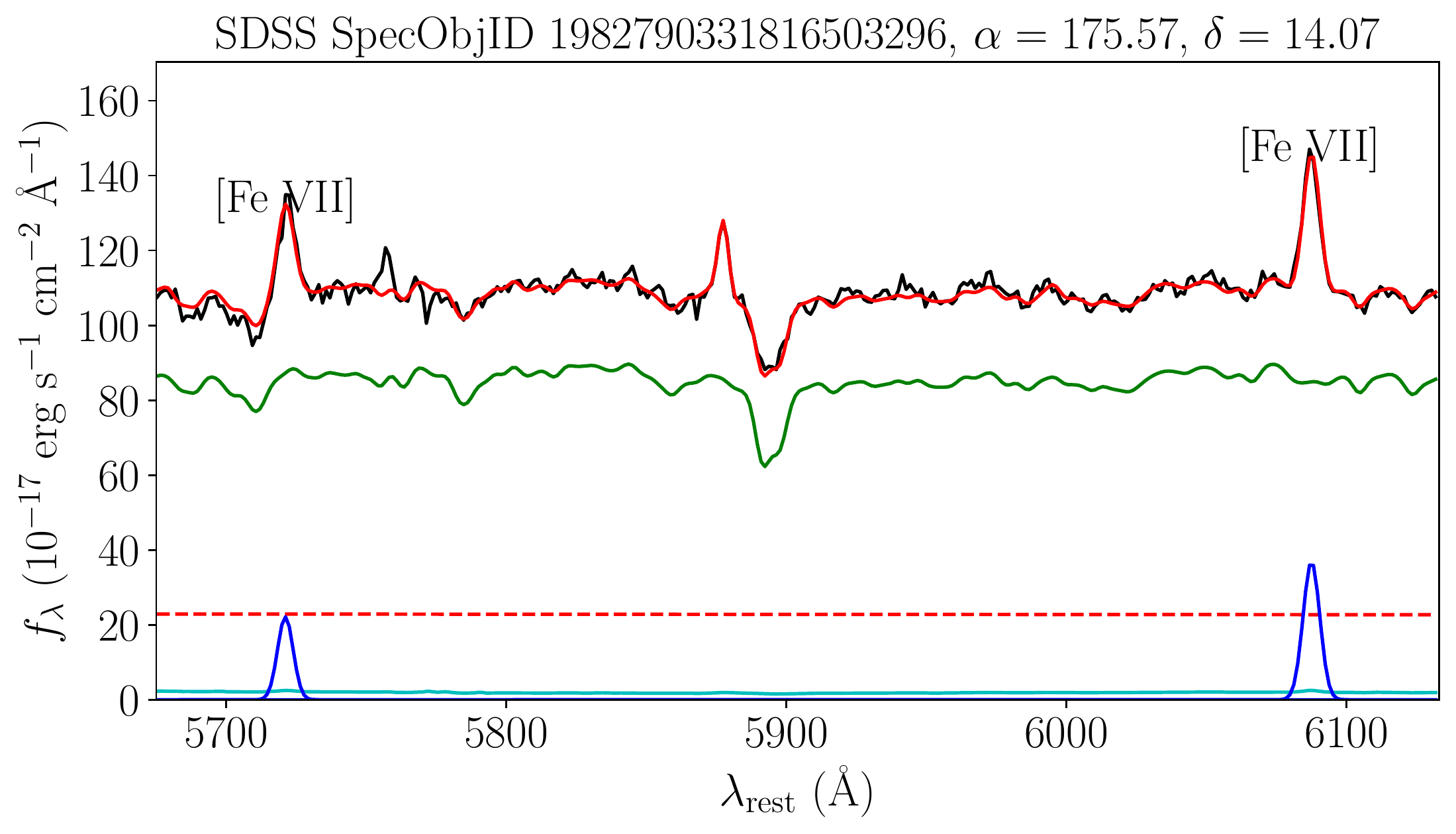}
    \includegraphics[width=\columnwidth]{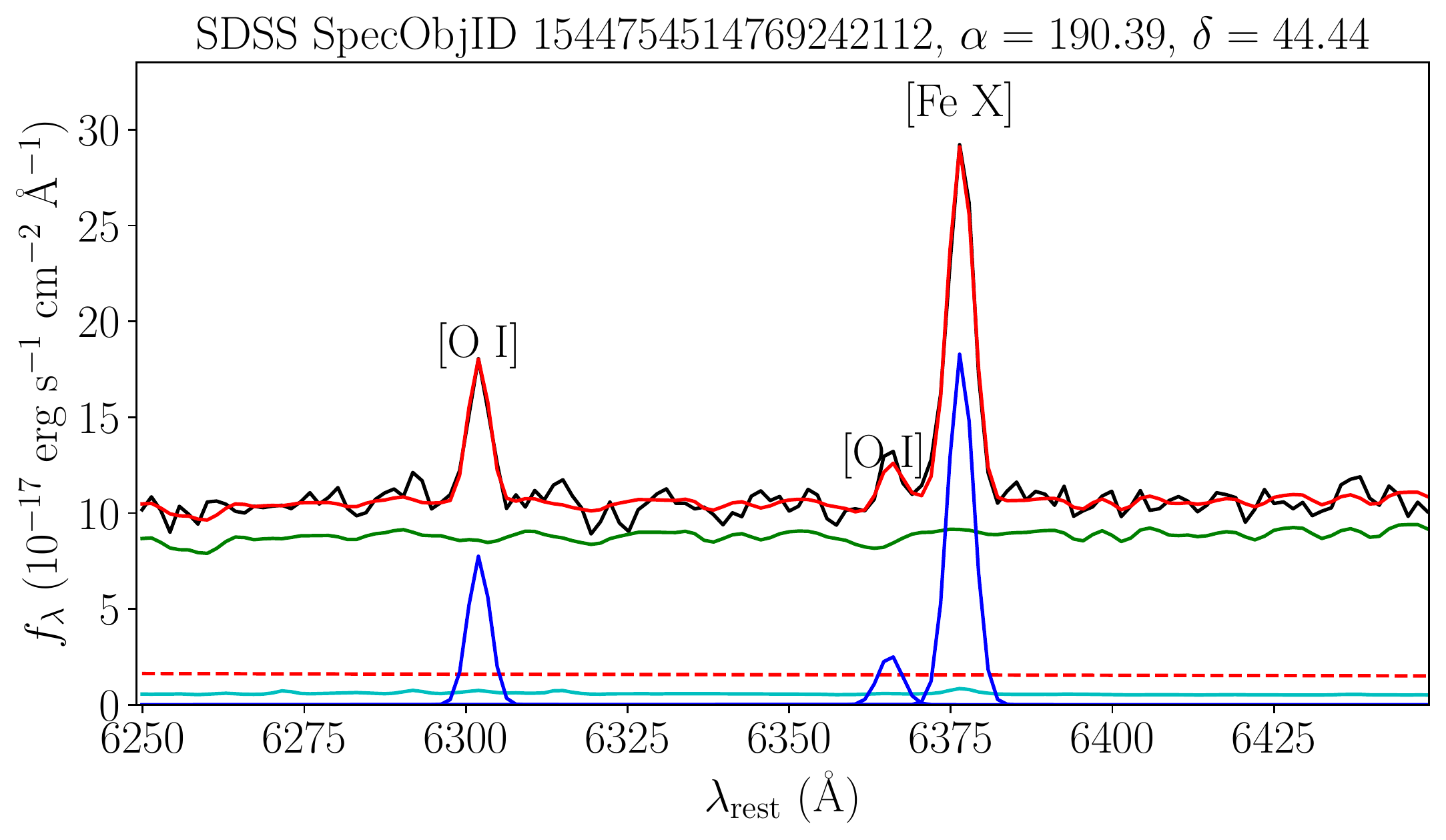}
    \includegraphics[width=\columnwidth]{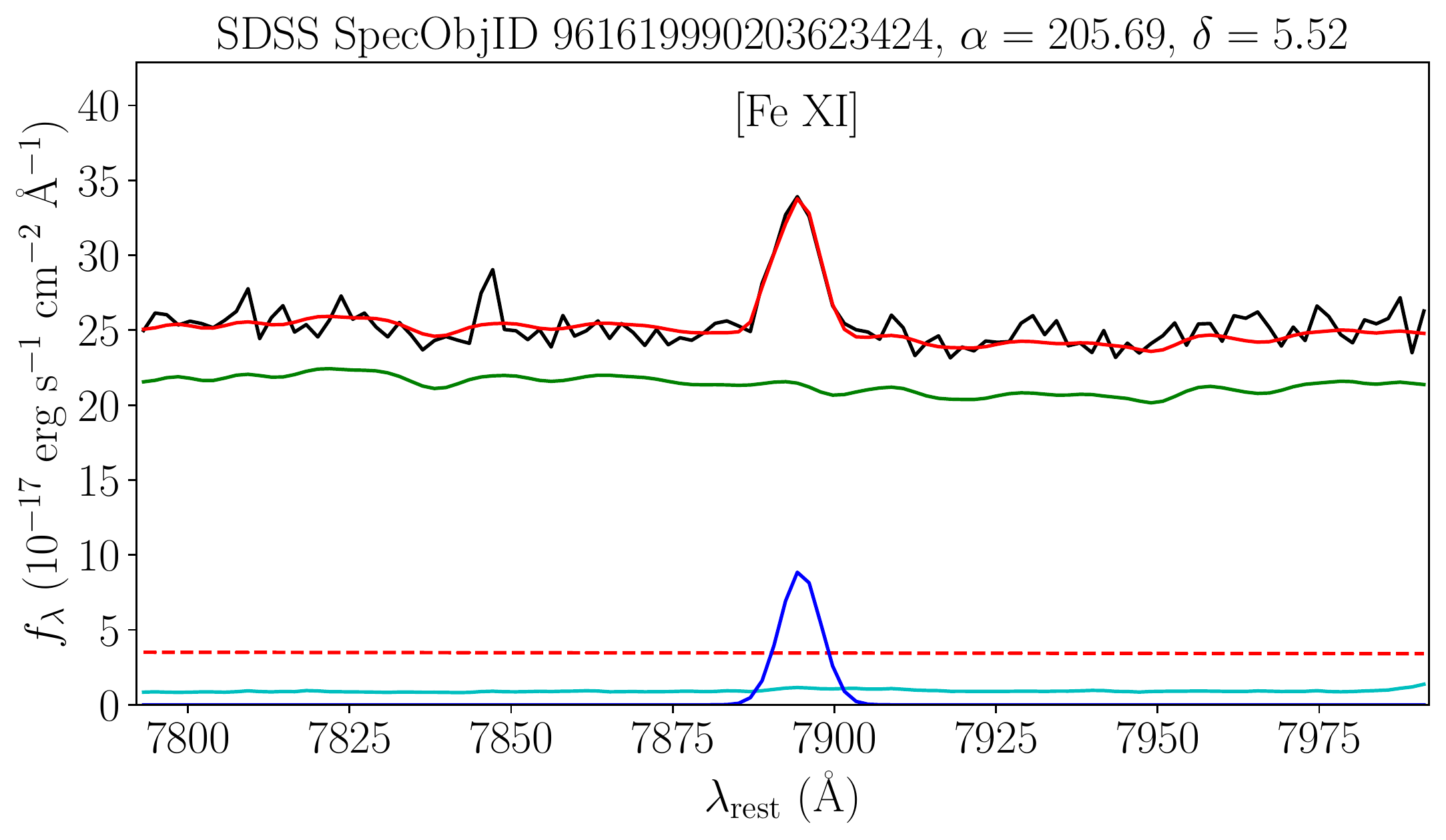}
    \caption{Examples of 6 individual spectra that exhibit emission of at least one coronal line in the optical.  The SDSS data, corrected for redshift and galactic extinction, are plotted in black, while the model and each of its components (AGN power law, host galaxy, and emission lines) are overlaid on the spectra.  Each coronal line is labeled, along with the (non-coronal) [\ion{O}{1}] line in the [\ion{Fe}{10}] spectrum, and the object's SDSS Spec Object ID and coordinates are labeled in each plot's title.}
    \label{fig:spectra}
\end{figure*}

The creation of the CLASS catalog begins with the SDSS MPA/JHU DR8 catalog \citep{2011AJ....142...72E,2011ApJS..193...29A,2006AJ....131.2332G,2013AJ....146...32S}\footnote{\href{https://wwwmpa.mpa-garching.mpg.de/SDSS/DR7/}{https://wwwmpa.mpa-garching.mpg.de/SDSS/DR7/}}, which contains 952,138 optical ($\sim$3300--8000 \AA) spectra of galaxies.  We choose this data release in particular to utilize the derived host galaxy properties available for each spectrum, including stellar masses and star formation rates.  We perform no initial pre-selection on this sample based on nuclear activity and instead conduct a search for all coronal line emission in the entire extragalactic population.  In this work, we systematically search for every coronal line that is present at optical wavelengths, from the [\ion{Ne}{5}] $\lambda\lambda$3346,3426 doublet at the blue end to the high ionization-potential [\ion{Fe}{11}] $\lambda$7892 at the red end. We note that most previous coronal line searches have exclusively targeted already identified AGNs, or are based on a single or small set of coronal lines. Moreover, previous studies often impose requirements on the detection or widths of other lines (e.g [\ion{O}{1}] $\lambda$6302 for [\ion{Fe}{10}] $\lambda$6374 searches) and use these as pre-selection strategies in generating the final sample. We emphasize that our methodology is designed to be inclusive, with no bias imposed in our pre-selection strategy. Our goal is to gain an understanding of the demographics of coronal line activity in the general galaxy population and to quantify for the first time their detection rates.  The full list of coronal lines can be found in Table \ref{tab:lines}.  

We construct a sample of candidate coronal line emitters by applying a pre-selection filtering algorithm to search for robust detections. 
% For each spectrum, we initially assess whether a coronal line is present using a highly simplified but optimized test that requires no fittings, likelihood maximizations, or time- and resource-intensive simulations.  
For each of the coronal lines listed in Table \ref{tab:lines}, we perform a linear fit to the continuum using two adjacent reference regions $\pm$30 \AA\ on either side of the the line of interest, adjusting the location as necessary to ensure that they fall on a flat and featureless part of the continuum, and subtract this from the average flux in a 20 \AA\ region centered on the coronal line. We then calculate the root-mean-square (RMS) deviation of the flux in the adjacent reference windows, which quantifies the noise in the continuum, and require that the average flux centered on each coronal line is at least $4\sigma$ above the noise. This filtering is achieved with a custom-made Python package called BIFR\"OST\footnote{\href{https://github.com/Michael-Reefe/bifrost}{https://github.com/Michael-Reefe/bifrost}}.  We note that this pre-selection algorithm is designed to identify the most robust detections in the entire sample, and will necessarily exclude marginal detections. 

To eliminate spurious detections and contamination from nearby sky lines, we impose the additional constraints that our coronal line candidate spectra must have at least 3 continuous pixels above $3\sigma$ and must not be within $\pm$20 \AA\ (or $\sim \pm 1000$ \kms) of the four most prominent sky lines at 5578.5, 5894.6, 6301.7, and 7246.0 \AA\ respectively.

The resulting sample of 2,834 candidate spectra
were then fit using the Bayesian AGN Decomposition Analysis for SDSS Spectra \citep[BADASS;][]{sexton_2020} software.  Spectra are fit with BADASS first using a likelihood maximization routine from SciPy, followed by a Markov Chain Monte Carlo (MCMC) fit using the affine-invariant Emcee sampler \citep{2013PASP..125..306F} to obtain robust uncertainties and covariances.  The BADASS fitting routine breaks down a spectrum into its base components, modeling the AGN power-law continuum, \ion{Fe}{2} emission, and spectral lines, simultaneously with the stellar line-of-sight velocity distribution (LOSVD) using the penalized pixel-fitting \citep[pPXF;][]{2017MNRAS.466..798C} method.  Each coronal line is modeled as a simple Gaussian profile with a free amplitude, velocity offset, and width.  We also do not fit the entire wavelength range of the spectrum at once, but instead subdivide it into smaller regions of a few hundred {\r a}ngstr\"oms to glean basic information about the continuum level. We require the model-fit flux of each line to be above $3\sigma$.

% small paragraph on how we fit specific lines - special case for [Fe X]
% small note that we didn't do full spectrum fitting, just with small regions to get an idea of the continuum level
For a few specific lines, we impose additional constraints on certain parameters.  Doublets such as [\ion{Fe}{7}] $\lambda\lambda$5720,6087 have their widths and offsets tied.  The flux ratio of the [\ion{O}{1}] $\lambda\lambda$6302,6365 doublet is fixed to 3, as predicted by atomic physics, and the line widths and velocity offsets are both tied to [\ion{N}{2}] $\lambda$6585, ensuring that the nearby [\ion{Fe}{10}] $\lambda$6374 can be robustly distinguished from [\ion{O}{1}].  

We then perform a final round of visual inspection before confirming each coronal line detection.  We present a selection of 6 individual spectra that show some of the most commonly detected coronal lines in Figure \ref{fig:spectra}, with the spectral fit overlaid. Our final sample contains 258 galaxies ($\sim$0.03\% of the catalog).  The coronal line luminosities in our sample range from $10^{34}$--$10^{42}$ \ergs, with a median value of $1.08 \times 10^{40}$ \ergs. The average and standard deviation of the luminosity for each coronal line is listed in Table \ref{tab:lines}.  The redshifts of our coronal line sample ranged from $0$--$0.3$, with an average of $0.105 \pm 0.100$. 
% For a more complete description of our sample selection, filtering algorithm, fitting process, and detailed information on line profiles and kinematics, see Reefe et al. (2022b, in prep.).
%BADASS models overlaid and split up by their base components---the host galaxy, AGN power law, and emission lines.

The galaxies in our sample are classified as AGNs or star forming (SF) galaxies based on optical narrow emission line ratios, using the Baldwin-Phillips-Terlevich (BPT) diagnostics \citep[][]{1981PASP...93....5B}, as well as mid-infrared color selection. For optical selection, we use the K03 \citep[][]{2003MNRAS.346.1055K} and K01 \citep[][]{2001ApJ...556..121K} selection criteria to classify spectra as AGNs or SF galaxies using fluxes provided in the MPA/JHU catalog, requiring a signal to noise $>3$ on the relevant narrow line detections. In order to classify galaxies in the sample according to their mid-infrared color, we matched the MPA/JHU DR8 catalog to the final public AllWISE catalog\footnote{\url{https://wise2.ipac.caltech.edu/docs/release/allwise/}},  where a galaxy is matched if the positions agree to within 6$''$. We require that all objects be detected with a signal-to-noise $>5$ in the 3.4 $\mu$m $W1$ and 4.6 $\mu$m $W2$ bands in order to employ a mid-infrared color diagnostic. We use the 3-band AGN color cut from \citet{2011ApJ...735..112J} to identify mid-infrared AGNs in the sample. In order to obtain information on the host galaxy morphologies, we use the magnitudes and Sersic indices, obtained from single component two dimensional fits to the surface brightness profiles, from the NASA-Sloan Atlas (NSA) catalog (version 1.0.1)\footnote{\href{https://www.sdss.org/dr13/manga/manga-target-selection/nsa/}{https://www.sdss.org/dr13/manga/manga-target-selection/nsa/}}. 
% Stellar masses and star formation rates (SFRs) are taking directly from the MPA/JHU DR8 catalog.

% Trying to rewrite the below paragraph:
%For the entire MPA/JHU catalog, we calculate intrinsic 1$\sigma$ errorbars on the flux around the regions that each coronal line is found by integrating a Gaussian distribution with amplitude equal to the noise of the region and width equal to the wavelength-dependent instrumental FWHM dispersion of the region.  We use these errorbars as a filter both for our detections and our nondetections by imposing that the average luminosity of each coronal line $\bar{L} \geqslant 3\sigma_L$, and throwing out any spectra where the error $\sigma_L$ does not satisfy this criterion.  This allows us to homogenize the control samples of each individual line, allowing their detection percentages to be directly comparable to each other.

One of our goals is to quantify the detection statistics for each of the 20 coronal lines in this study. Because the luminosity of the various coronal lines varies considerably, as does the signal-to-noise of the observations, \rev{the total number of spectra employed in the calculated detection fraction varies from line to line. In particular, in calculating detection fractions, we exclude from consideration spectra in which the sensitivity is not sufficient to enable a 3$\sigma$ detection of the given coronal line with a luminosity equal to the average luminosity amongst our detections. In addition, in calculating each detection fraction, we exclude from consideration spectra in which the given coronal line is redshifted out of the SDSS spectral range. We calculate the sensitivity of each spectrum by integrating a Gaussian with amplitude equal to the RMS of the flux and width equal to the instrumental FWHM dispersion, both calculated at the location of the coronal line.}  This allows us to account for the non-uniform sensitivity of the observations to each of the coronal lines, allowing their detection percentages to be directly comparable to each other, given the sensitivity of SDSS.
% in calculating a detection fraction, we only include observations in which 1) the coronal line is not redshifted out of the SDSS spectral range, and 2) the sensitivity of the observation is sufficient to detect a coronal line with luminosity equal to the average value for the given line amongst the detections in the sample. This allows us to account for the non-uniform sensitivity of the observations to each of the coronal lines, allowing their detection percentages to be directly comparable to each other, given the sensitivity of SDSS.

\section{Results}
\subsection{Detection Statistics and Comparison with other Diagnostics}

\input{line_table_letter}

We find that coronal line emission is extremely rare in SDSS spectra of galaxies. Given the total sample of 258 galaxies that display at least one coronal line identified through our pipeline, the detection fraction using binomial statistics, with 68\% confidence bounds, is $0.0271^{+0.0018}_{-0.0017}$\%, significantly lower than the fraction obtained using other commonly employed diagnostics for nuclear activity. In comparison, the detection fraction of galaxies identified as K01 AGNs  and WISE AGNs is $16.821 \pm 0.031$\% and $2.002^{+0.012}_{-0.011}$\%, respectively. The detection fraction and luminosity distributions vary considerably for the various lines, as can be seen from Table \ref{tab:lines}.  Among the most commonly found coronal lines are [\ion{Fe}{10}] $\lambda$6374, the [\ion{Fe}{7}] $\lambda\lambda$6087,5720 doublet, and the [\ion{Ne}{5}] $\lambda\lambda$3346,3426 doublet, with detection fractions upwards of 0.0407\%.  The least common, on the other hand, are [\ion{S}{12}] $\lambda$7609, and [\ion{Fe}{5}] $\lambda$3891 with only 2 detections.  \rev{The maximum number of coronal lines that we detect together in a single galaxy is only 6, with two such galaxies, SDSS specObjIDs 2411797517822879744 and 2904951020922628096.  The former exhibits [\ion{Fe}{11}] $\lambda$7892, [\ion{Fe}{10}] $\lambda$6374, [\ion{Fe}{7}] $\lambda\lambda$6087,5720, [\ion{Ca}{5}] $\lambda$5309, and [\ion{Fe}{7}] $\lambda$5159.  The latter exhibits almost the same set of lines, but shows [\ion{Fe}{14}] $\lambda$5303 instead of [\ion{Fe}{7}] $\lambda$5159.  It is clear from this study that it is rare to detect many coronal lines in a single spectrum; 49\% of the coronal line emitting galaxies have only one coronal line detected. We note that because our filtering algorithm is designed to pick out robust detections and therefore may exclude marginal detections, our catalog does not include all coronal line detections reported previously in the literature \citep[e.g.][]{10.1111/j.1365-2966.2009.14961.x,2021ApJ...922..155M}. We examine the overlap between the various samples in more detail in Reefe et al. 2022b.}

\rev{The most luminous coronal line detected in the sample is [\ion{Ne}{5}] $\lambda$3426, with an average luminosity of $1.62 \times 10^{41}$ erg s$^{-1}$. The strength of this line, in conjunction with its low IP and its location in a relatively flat part of the spectrum is consistent with the [\ion{Ne}{5}] $\lambda\lambda$3426,3346 doublet having one of the highest detection fractions seen in the optical. The next most luminous line is the [\ion{Fe}{5}] $\lambda$3839, with an average luminosity of $1.58 \times 10^{41}$ erg s$^{-1}$. We note that the [\ion{Fe}{5}] doublet is located nearly directly on top of two higher-order Balmer lines, H$\zeta$ at 3889 \AA\ and H$\eta$ at 3835 \AA, making the detection of the line more challenging.  Within the few detections for [\ion{Fe}{5}] that we claim, the lower-order H$\varepsilon$ is not visible in the spectra, so it is unlikely our results are contaminated by these Balmer lines.  
%Lines between $\sim$4400--5500 \AA\ will also be more difficult to detect in type I AGN spectra, as opposed to the sample used for this survey, due to \ion{Fe}{2} contamination (see Doan et al. 2022, in prep.). 
The least luminous coronal line in the sample is the [\ion{Fe}{7}] $\lambda$5276, with an average luminosity of $1.20 \times 10^{37}$ erg s$^{-1}$, potentially making it the most difficult line to detect in the optical.}

\begin{figure*}
    \centering
    \includegraphics[width=.9\textwidth]{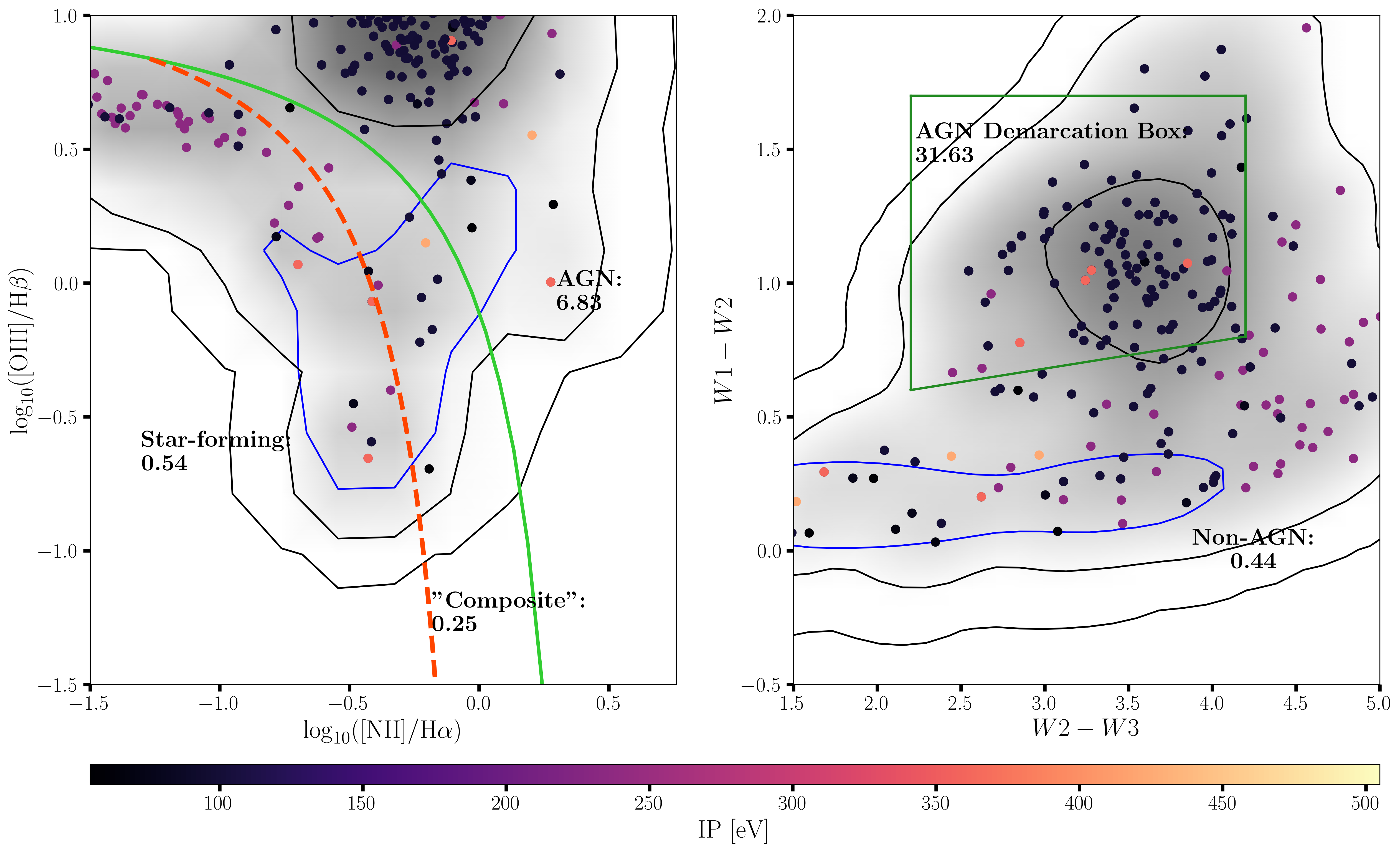}
    \caption{\textit{Left}: A BPT ratio plot of all the galaxies in our subsample that have BPT information.  The color of each point corresponds to the ionization potential of the most energetic coronal line detection in the galaxy.  The star-forming, ``composite,'' and AGN regions are marked with text and separated by the curves defined in \citet{2003MNRAS.346.1055K} and \citet{2001ApJ...556..121K}.  The black contours and shading highlight the probability density of the subsample, while the blue contour shows the 68\% threshold of the entire MPA/JHU DR8 catalog.  Each region is also annotated under its label with the excess of coronal line detections in that region, defined as the fraction of coronal line detections over the total number of spectra in that region, normalized by the overall coronal line detection fraction (0.0271\%). \\
    \textit{Right}: A WISE color-color plot of all the galaxies in our subsample that have WISE observations.  The coloring of the points and contours is defined the same way as the BPT plot.  The AGN demarcation box is defined from \citet{2011ApJ...735..112J}.}
    \label{fig:bpt_wise}
\end{figure*}

In Figure \ref{fig:bpt_wise}, we show the narrow line ratios employed in the Baldwin-Phillips-Terlevich (BPT) diagram \citep{1981PASP...93....5B} and WISE \citep{2010AJ....140.1868W} colors of our coronal line sample in comparison to the entire catalog, together with the standard AGN demarcation lines employed in the literature. The highest ionization potential of the coronal lines detected in a galaxy is indicated by the color bar. While a large fraction of our coronal line emitters do reside in the K01 region (72\%), and mid-infrared AGN region (59\%), a significant fraction of coronal line emitters (11\%) are not identified as AGNs using optical narrow line ratios or mid-infrared color selection.  Moreover, a key and striking result of this work is that the majority of [\ion{Fe}{10}] and higher IP detections are found in the star-forming region of the BPT plot and outside the Jarrett AGN box in the WISE plot. Note that although AGNs are expected to produce a hard radiation field, coronal line emission is not universally seen in either BPT-selected or mid-infrared selected AGNs, a result that is consistent with numerous other studies \citep[e.g.,][]{2000AJ....119.2605N,2006A&A...457...61R,2017MNRAS.467..540L}. Only $\sim 0.2$\% and $\sim 0.9$\% of BPT and mid-infrared AGNs showed coronal line emission detectable by SDSS in our sample. Given the average luminosity of coronal lines that are detected in our sample, and considering only those observations with the sensitivity to detect them,  only $0.2$\% of K01 and $6.1$\%, of WISE AGNs display coronal line emission identified in our pipeline. Luminous optical coronal line emission is therefore intrinsically rare, a result that is explored in McKaig et al. (2022, in prep.).

\subsection{Host galaxy properties}

\begin{figure}
    \centering
    \includegraphics[width=\columnwidth]{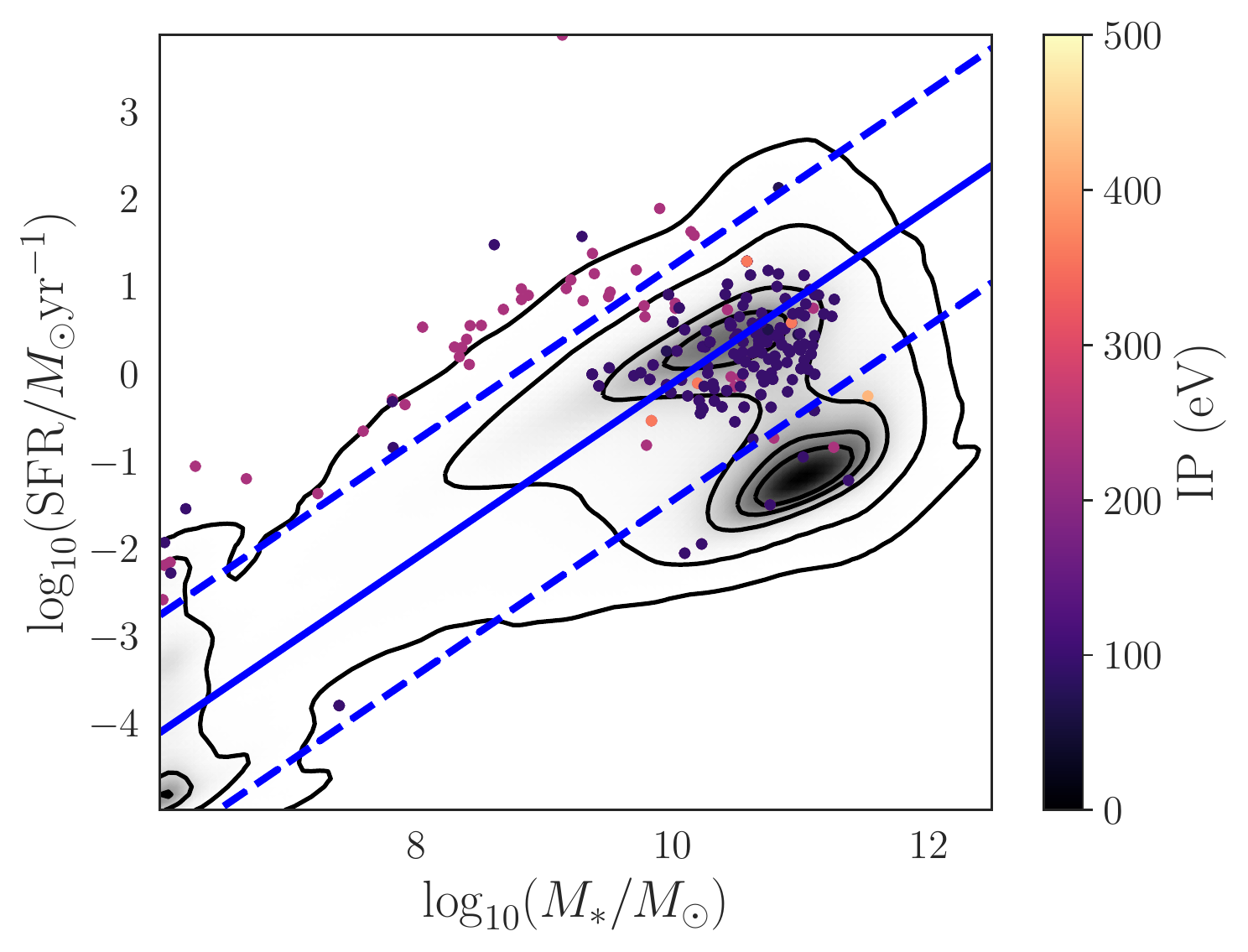}
    \caption{Galaxy star formation rate vs. stellar mass.  A Gaussian kernel density estimation of the full MPA/JHU catalog is plotted in grey, while our coronal line detections are plotted individually, colored by the ionization potential.  The main sequence relationship from \citet{2017ApJ...851...22M} is plotted in blue, and the $1\sigma$ standard deviation of the full sample is shown in the dashed blue lines.}
    \label{fig:sfr_mass}
\end{figure}

%Further examining the star formation rates of galaxies that exhibit coronal lines provides supporting evidence for the trend hinted at by the BPT plot.  
Having identified coronal line emitters in SDSS, we now explore the host galaxy demographics of this population and compare it to those of other galaxy subclasses. Specifically, we explore the SFRs, stellar masses, and galaxy morphologies of the sample. We use the  the [\ion{N}{2}]/H$\alpha$ ratio, the most robust metallicity indicator in galaxies where the primary source of ionization is not well-constrained \citep[e.g.,][]{2006NewAR..50..743G} to gain insight into the metallicity of the host galaxies of our sample in comparison to other subclasses. In comparing host galaxy morphology, we use the S\'ersic index, $n_{\rm sersic}$, with low values indicating disk-like morphologies and higher values corresponding to bulge-dominated morphologies. Because the S\'ersic fits depend on the spatial extent and hence distance to the galaxies, we match by $r$-band magnitude when comparing S\'ersic indices in BPT-AGNs with coronal line emitters.  In Figure \ref{fig:sfr_mass}, we plot the SFR against the stellar mass of our sample with the main sequence relationship from \citet{2017ApJ...851...22M} overlaid. The highest ionization potential coronal line detected in a given galaxy is indicated by the color bar. As can be seen, a significant fraction of the coronal line detections (50 galaxies, or 19.4\% of the 258 coronal line emitters) lie more than $1\sigma$ away from the main sequence relationship, with the  vast majority of the higher ionization potential coronal line detections preferentially found well above the relationship.  Similar results are found by \citet{2021ApJ...922..155M} in their sample of [\ion{Fe}{10}]-emitting dwarf galaxies. While it is clear that coronal line emission is preferentially seen in galaxies with elevated SFRs, this does not necessarily imply that the emission lines are generated by stellar processes, as pointed out by \citet{2021ApJ...922..155M}. Elevated accretion activity onto a SMBH or an elevated tidal-disruption event (TDE) rate may be correlated with elevated SFRs in galaxies, as nuclear gas availability may provide the fuel for both accretion and star formation activity. There is no correlation seen between the host galaxy specific SFR and coronal line luminosity, a result we explore further in Reefe et al. (2022b., in prep.)
%If we briefly consider the potential of coronal line emission as a diagnosis for AGN-like activity, with higher ionization potential lines hinting at hotter, more energetic nuclei, then these would also be correlated with higher star formation rates, potentially triggered by hot gas and dust being ejected from the galactic nucleus and into the disk, providing a constantly replenishing source of new material for birthing new stars.

\begin{figure*}
    \centering
    \includegraphics[width=\columnwidth]{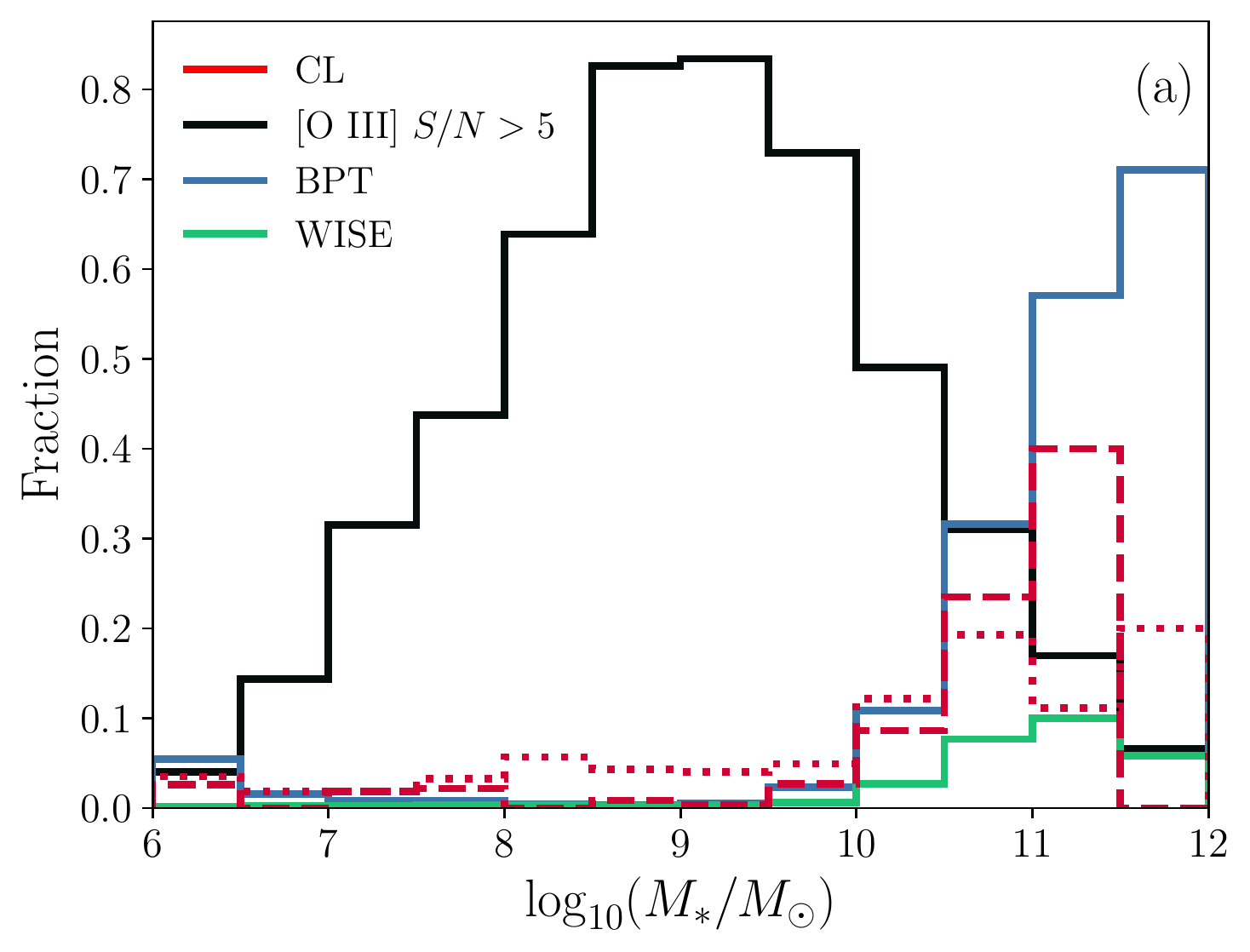}
    \includegraphics[width=\columnwidth]{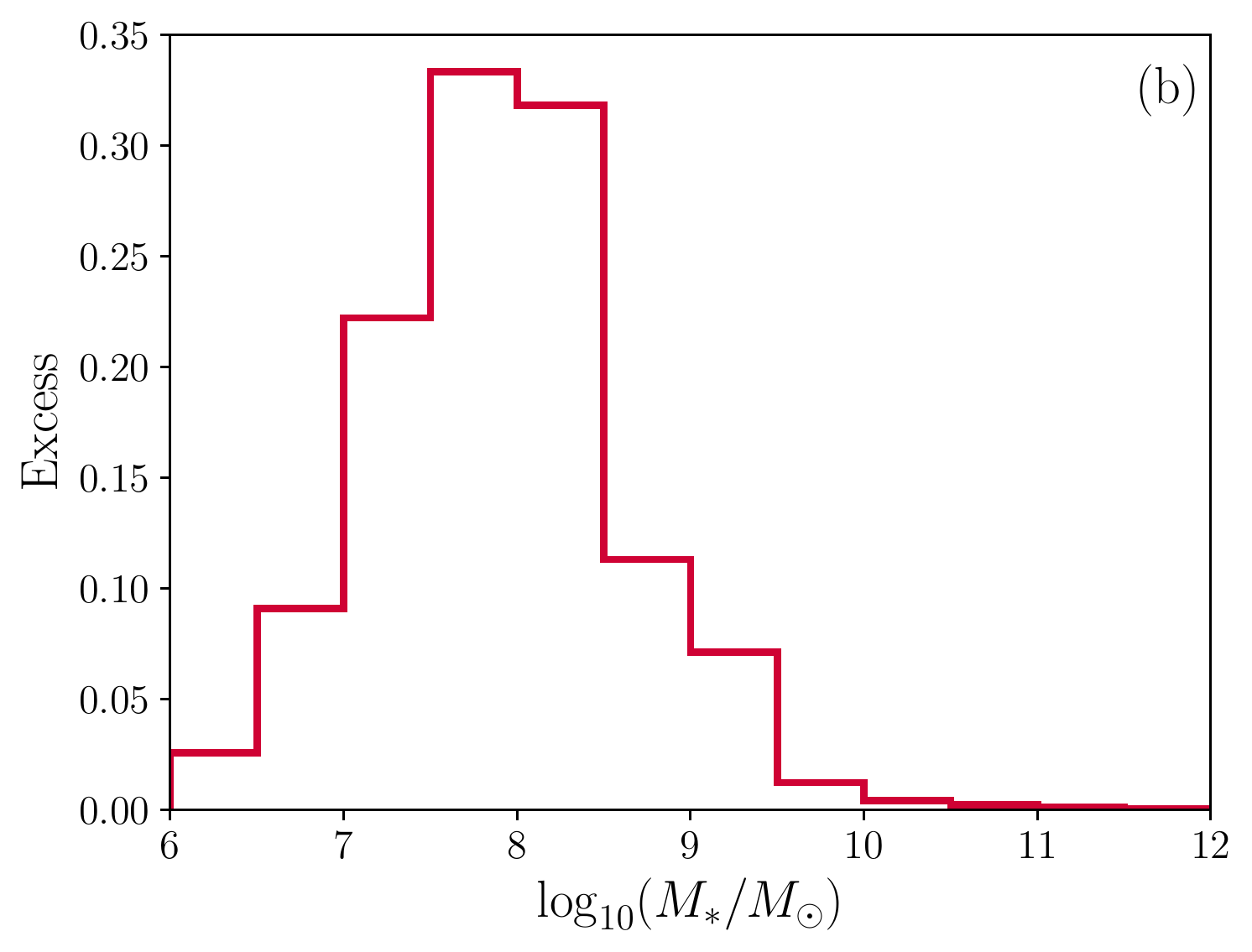}
    \includegraphics[width=\columnwidth]{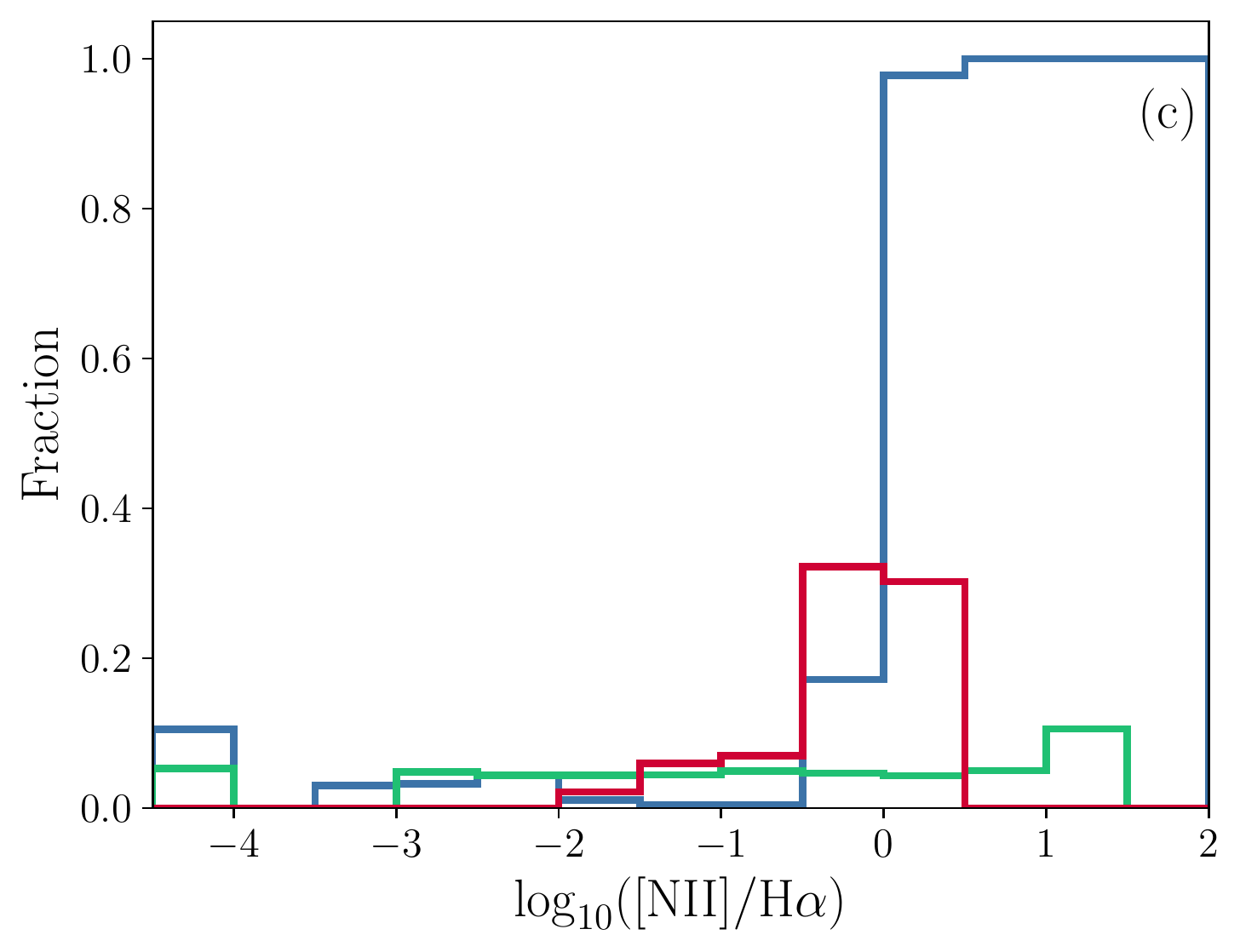}
    \includegraphics[width=\columnwidth]{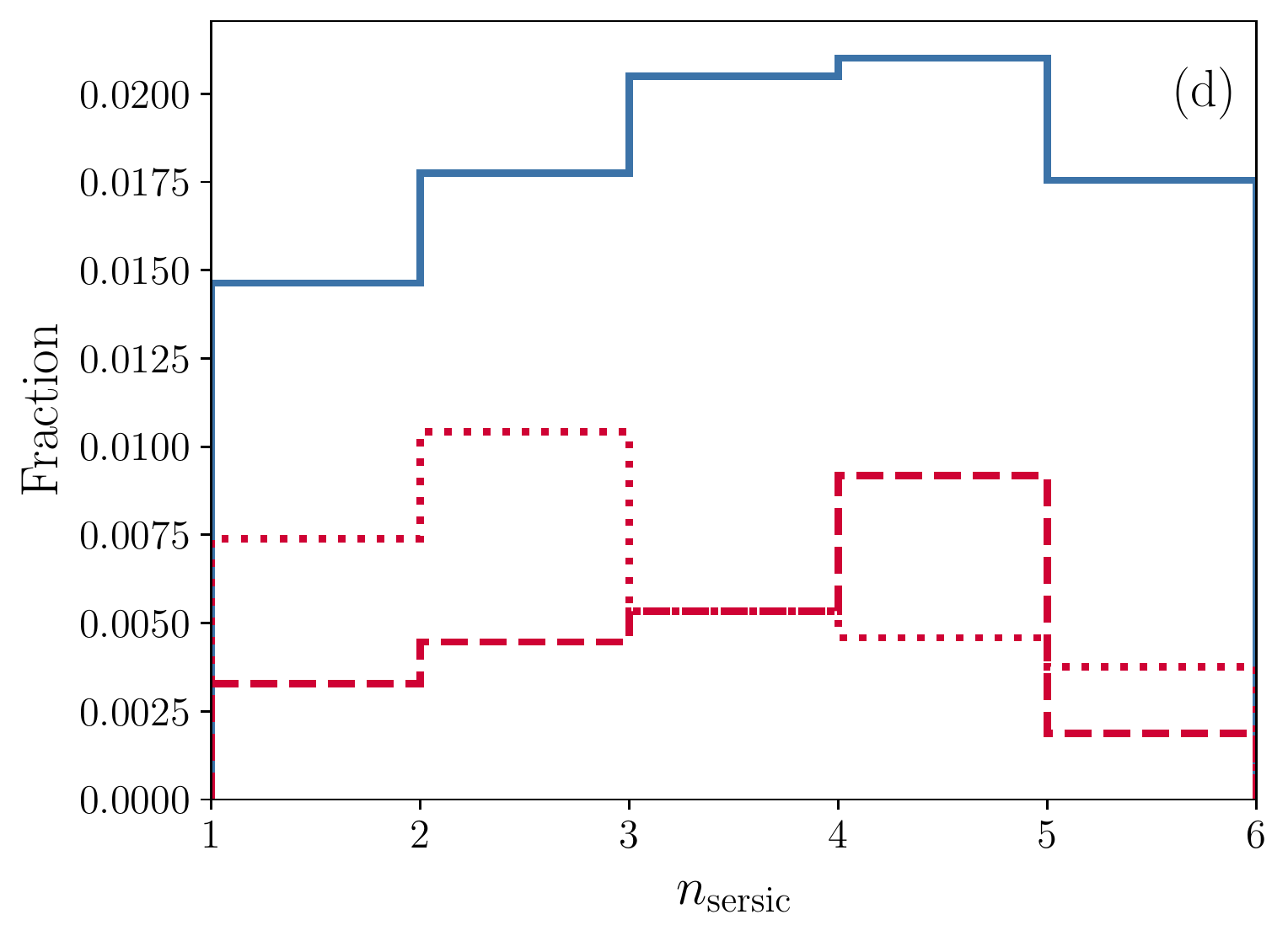}
    \caption{Histograms of various subsets of our galaxy sample, with stellar mass, metallicity, and Sersic index on the abscissae.  The ordinate is the fraction of galaxies that meet certain criteria within each bin.  The solid black line is the fraction of \rev{the full galaxy sample of 952,138} that have an [\ion{O}{3}] $S/N \geqslant 5$, the solid blue line is the fraction of those galaxies \rev{that meet the [\ion{O}{3}] $S/N$ requirement (of which there are 354,905)} that are classified as AGN via their BPT ratios \rev{using the \citet{2001ApJ...556..121K} criterion,} and the solid green line is the fraction of \rev{the [\ion{O}{3}] galaxies that are classified as} AGNs by their WISE colors as in \citet{2011ApJ...735..112J}.  The red lines indicate the fraction \rev{of} coronal line emitters \rev{out of the initial subsample of 2,834 candidates marked by the BIFR\"OST analysis which also meet the [\ion{O}{3}] $S/N$ threshold, of which there are 1,651.} The solid line \rev{includes} all coronal line emitters, the dashed line \rev{includes} only those with IPs $<$ 200 eV, and the dotted line \rev{includes} only those with IPs $>$ 200 eV.  In panel b, we plot the excess of coronal line detections as a fraction of BPT-identified AGNs \rev{from panel a, of which there are 62,410}.  In panel d, the data are subsampled such that the distributions in $r$ magnitude matches between the control BPT AGNs and the sample coronal line emitters\rev{--the BPT fraction is again out of the full sample, while the coronanl line fraction is out of the $r$-magnitude-matched BPT subsample, of which there are 5,100}.}
    \label{fig:histograms}
\end{figure*}

In Figure \ref{fig:histograms}, we explore the host galaxy demographics of coronal line emitters relative to those of BPT-selected and WISE AGNs. In these plots, we only consider emission line galaxies in which the [\ion{O}{3}] line is detected with $S/N \geqslant 5$, \rev{which constitutes $\sim 1/3$ of the original sample,} and plot the fraction of galaxies in various subclasses of nuclear activity. We replicate the well-known trend that optical narrow line diagnostics select AGNs in more massive (Figure \ref{fig:histograms}a), metal rich (Figure \ref{fig:histograms}c), and bulge-dominated (Figure \ref{fig:histograms}d) hosts \citep[see Figure 5 in][]{2003MNRAS.346.1055K}. In contrast, we do not see the same dramatic drop in coronal line emitters and WISE AGNs in lower mass galaxies, despite the fact that the lower mass galaxies suffer greater flux incompleteness. Moreover, a key and striking result from this work is that there is a dramatic increase with decreasing stellar mass in coronal line emitters that are not identified as BPT AGNs, as can be seen by Figure \ref{fig:histograms}b. We also see in Figure \ref{fig:histograms}a that in lower mass hosts, the higher ionization potential coronal lines are more prevalent than the lower ionization potential lines, in contrast to what is found in the higher mass galaxies. Coronal line emitters are also found in lower metallicity galaxies based on their [\ion{N}{2}]/H$\alpha$ ratios, as can be seen in Figure \ref{fig:histograms}c. Higher ionization potential coronal lines also possibly favor disk-like hosts compared with lower ionization potential lines and BPT-selected AGN hosts (Figure \ref{fig:histograms}d). While the limited sample size of coronal line emitters precludes us from conducting a statistically rigorous analysis, this study clearly demonstrates that coronal line emission is uncovering a different host galaxy population than is revealed by traditional optical narrow line ratios.

% increase the line widths and font size

\section{Discussion and Conclusions}
We have constructed the CLASS survey, with the goal of providing for the first time a complete census of all optical coronal line emission in the galaxy population observed by SDSS. Our survey is designed to pick out all robust detections of the 20 optical coronal lines accessible in the SDSS wavelength regime; marginal detections are not included in the current survey. We find that optical coronal line emission is extremely rare in the galaxy population, even in AGNs identified through optical narrow line ratios or mid-infrared color selection, and even when accounting for the sensitivity of SDSS. Our detection rate in AGNs is significantly lower than that seen in the near-infrared spectra of AGNs \citep{2011ApJ...743..100R,2017MNRAS.467..540L,2018ApJ...858...48M}, possibly due to dust extinction, a result that underscores the crucial need for JWST in future coronal line studies. This study also demonstrates that, while rare, a significant fraction of coronal line emitters are found in galaxies with no signs of nuclear activity based on their optical narrow line ratios or mid-infrared color, with the majority of coronal line emitters with no optical or mid-IR AGN indication being found in lower mass hosts. Coronal line emitters are preferentially found in galaxies with elevated star formation rates, and tend to reside in more metal-poor, and later-type galaxies compared to BPT AGNs.

In this work, we have not explored the origin of the coronal line emission, but instead our goal is to provide a complete census of coronal line activity in SDSS and explore their host galaxy properties. Optical coronal line emission has been observed in extragalactic supernovae \citep{1990AJ....100.1588B,2009ApJ...695.1334S, 2009ApJ...707.1560I}, and can be produced by Wolf-Rayet stars, planetary nebula or shocks from starburst-driven winds \citep{1999A&A...345L..17S,2008ApJ...678..686A,2008ApJS..178...20A,2012ApJ...754...28Z}, but their luminosities are generally significantly lower than the luminosities of the coronal lines presented in this work. Photoionization from stars in lower metallicity galaxies can also potentially produce coronal lines. Lower metallicity galaxies are dominated by more massive stars which produce a harder radiation field \citep{1986MNRAS.223..811C}, which in turn can produce an emission line spectrum dominated by higher ionization potential lines, a result that has been observed with the detection of [\ion{Ne}{5}] $\lambda\lambda$3346,3426 in blue compact dwarf galaxies \citep{2012MNRAS.427.1229I, 2021MNRAS.508.2556I}. However, we reiterate that the luminosities of the coronal lines in this sample are generally larger than those produced by purely stellar processes, ranging from 1--2 orders of magnitude above those typically seen from stellar processes. Finally, coronal line emission can be produced by TDEs or accretion onto a SMBH \citep{2012ApJ...749..115W}. If this is the case, the prevalence of coronal line emission in dwarf galaxies in our sample, with the lines with highest ionization potential preferentially found in lower mass hosts, is consistent with the theory that lower mass hosts with lower mass black holes produce a hotter accretion disk, which in turn enhances the high ionization coronal line spectrum \citep{2018ApJ...861..142C} and causes the optical narrow line ratios employed in the BPT diagram to be indistinguishable from those of star forming galaxies \citep{2019ApJ...870L...2C}. In Reefe et al. (2022b, in prep.), we present an in-depth description of the CLASS catalog, provide the full list of line fluxes and kinematics, and explore the origin of the coronal line emission in our sample. Given the sensitivity of SDSS and the faintness of coronal line emission, more sensitive surveys are critical to unlock their full diagnostic potential. \rev{Infrared coronal lines are significantly stronger than optical coronal lines. For example, in the nearby AGN, Circinus, the infrared coronal lines are up to over two magnitudes more luminous than the optical coronal lines \citep{1994A&A...288..457O,1996A&A...315L.109M}, with an even more signifcant enhancement of the infrared coronal lines relative to the optical coronal lines anticipated for intermediate mass black holes \citep[IMBHs,][]{2018ApJ...861..142C}.} Follow-up observations with JWST may \rev{thus be} a powerful avenue for uncovering \rev{IMBHs} and constraining their properties.       

\section{Acknowledgements}

J.M.C. would like to acknowledge a NASA Postdoctoral Program (NPP) fellowship at Goddard Space Flight Center, administered by ORAU through contract with NASA.

% We thank the referee for their thoughtful comments. 
The MCMC simulations carried out in this work were run on ARGO and HOPPER, research computing clusters provided by the Office of Research Computing at George Mason University, VA. (\url{ http://orc.gmu.edu}).

This research made use of Astropy,\footnote{\url{http://www.astropy.org}} a community-developed core Python package for Astronomy \citep{2013A&A...558A..33A}, as well as \textsc{topcat} \citep{2005ASPC..347...29T}.  
% J.M.C. gratefully acknowledges support from the National Science Foundation Graduate Research Fellowship Program.

% something for the SDSS?
Funding for SDSS-III has been provided by the Alfred P. Sloan Foundation, the Participating Institutions, the National Science Foundation, and the U.S. Department of Energy Office of Science. The SDSS-III web site is \href{http://www.sdss3.org/}{http://www.sdss3.org/}.

SDSS-III is managed by the Astrophysical Research Consortium for the Participating Institutions of the SDSS-III Collaboration including the University of Arizona, the Brazilian Participation Group, Brookhaven National Laboratory, Carnegie Mellon University, University of Florida, the French Participation Group, the German Participation Group, Harvard University, the Instituto de Astrofisica de Canarias, the Michigan State/Notre Dame/JINA Participation Group, Johns Hopkins University, Lawrence Berkeley National Laboratory, Max Planck Institute for Astrophysics, Max Planck Institute for Extraterrestrial Physics, New Mexico State University, New York University, Ohio State University, Pennsylvania State University, University of Portsmouth, Princeton University, the Spanish Participation Group, University of Tokyo, University of Utah, Vanderbilt University, University of Virginia, University of Washington, and Yale University. 

This publication makes use of data products from the Wide-field Infrared Survey Explorer, which is a joint project of the University of California, Los Angeles, the Jet Propulsion Laboratory/California Institute of Technology, and NEOWISE, which is a project of the Jet Propulsion Laboratory/California Institute of Technology. WISE and NEOWISE are funded by the National Aeronautics and Space Administration.

\facilities{Sloan, WISE}

\software{
\texttt{astropy} \citep{2013A&A...558A..33A}, 
\texttt{numpy},\citep{2020Natur.585..357H}
\texttt{scipy},\citep{2020NatMe..17..261V}
\texttt{emcee} \citep{2013PASP..125..306F},
\texttt{pPXF} \citep{2017MNRAS.466..798C},
\textsc{topcat} \citep{2005ASPC..347...29T},
\textsc{badass} \citep{sexton_2020},
\texttt{bifrost} (\href{https://github.com/Michael-Reefe/bifrost}{https://github.com/Michael-Reefe/bifrost})
}

\nocite{1988AJ.....95...45A}
\nocite{2018ApJ...861..142C}

\bibliographystyle{yahapj}
\bibliography{main}

% \begin{thebibliography}{}
%\newpage
% \providecommand\natexlab[1]{#1}
% \providecommand\JournalTitle[1]{#1}

% \bibitem[Appenzeller \& Oestreicher(1988)]{appenzeller1988} Appenzeller, I. \& Oestreicher, R.\ 1988, \aj, 95, 45. doi:10.1086/114611

% \bibitem[Cann et al.(2018)]{cann2018} Cann, J.~M., Satyapal, S., Abel, N.~P., et al.\ 2018, \apj, 861, 142. doi:10.3847/1538-4357/aac64a

%\input{./bibliography}

% \end{thebibliography}

%\appendix
%\section{appendix section}

\end{document}

%% file: line_table_letter.tex
\startlongtable
\begin{deluxetable*}{lccccccc}
\tabletypesize{\footnotesize}
\tablecaption{Optical coronal emission lines from 3346--7892 \AA. For the luminosity, the values shown are (mean) $\pm$ (standard deviation). The detection confidence levels are 68\%.}
\tablehead{\colhead{Line} & \colhead{Wavelength}\textsuperscript{1} & \colhead{Critical Density} & \colhead{Ionization Potential}\textsuperscript{2} & \colhead{Transition} &\colhead{Luminosity} & \colhead{Detections} & \colhead{Number} \\
\colhead{} & \colhead{(\AA)} & \colhead{(cm$^{-3}$)} & \colhead{(eV)} & \colhead{} & \colhead{$\log L/$erg s$^{-1}$ (dex)} & \colhead{(\%)} & \colhead{}}
\decimals
\startdata
\lbrack \ion{Fe}{11}\rbrack & 7891.800 & $6.39 \times 10^8$ & 262.10 & $^3$P$_2 - ^3$P$_1$ & $38.92 \pm 1.27$ & ${0.0064}^{+0.0029}_{-0.0021}$ & 9 \\ 
\lbrack \ion{S}{12}\rbrack & 7611.000 & $7.09 \times 10^9$ & 504.78 & $^2$P$^0_{1/2} - ^2$P$^0_{3/2}$ & $39.40 \pm 0.28$ & ${0.00058}^{+0.00076}_{-0.00037}$ & 2 \\ 
\lbrack \ion{Fe}{10}\rbrack & 6374.510 & $4.45 \times 10^8$ & 235.04 & $^2$P$^0_{3/2} - ^2$P$^0_{1/2}$ & $39.00 \pm 1.05$ & ${0.0407}^{+0.0043}_{-0.0039}$ & 107 \\ 
\lbrack \ion{Fe}{7}\rbrack & 6087.000 & $4.46 \times 10^7$ & 99.00 & $^3$F$_3 - ^1$D$_2$ & $39.96 \pm 0.59$ & ${0.0146}^{+0.0015}_{-0.0014}$ & 111 \\ 
\lbrack \ion{Fe}{7}\rbrack & 5720.700 & $3.72 \times 10^7$ & 99.00 & $^3$F$_2 - ^1$D$_2$ & $39.77 \pm 0.59$ & ${0.0158}^{+0.0016}_{-0.0015}$ & 114 \\ 
\lbrack \ion{Ar}{10}\rbrack & 5533.265 & $2.36 \times 10^9$ & 422.60 & $^2$P$^0_{3/2} - ^2$P$^0_{1/2}$ & $38.83 \pm 2.83$ & ${0.0028}^{+0.0019}_{-0.0012}$ & 5 \\ 
\lbrack \ion{Fe}{6}\rbrack & 5335.180 & $6.32 \times 10^6$ & 75.00 & $^4$F$_{3/2} - ^4$P$_{1/2}$ & $39.74 \pm 2.14$ & ${0.00131}^{+0.00064}_{-0.00045}$ & 8 \\ 
\lbrack \ion{Ca}{5}\rbrack & 5309.110 & $6.63 \times 10^7$ & 67.10 & $^3$P$_2 - ^1$D$_2$ & $40.11 \pm 0.58$ & ${0.00137}^{+0.00055}_{-0.00040}$ & 11 \\ 
\lbrack \ion{Fe}{14}\rbrack & 5302.860 & $3.99 \times 10^8$ & 361.00 & $^2$P$^0_{1/2} - ^2$P$^0_{3/2}$ & $39.94 \pm 0.66$ & ${0.00095}^{+0.00051}_{-0.00035}$ & 7 \\ 
\lbrack \ion{Fe}{7}\rbrack & 5276.380 & $2.98 \times 10^6$ & 99.00 & $^3$F$_4 - ^3$P$_2$ & $37.08 \pm 1.76$ & ${0.0057}^{+0.0026}_{-0.0018}$ & 9 \\ 
\lbrack \ion{Fe}{6}\rbrack & 5176.040 & $3.29 \times 10^7$ & 75.00 & $^4$F$_{9/2} - ^2$G$_{9/2}$ & $39.47 \pm 2.55$ & ${0.00070}^{+0.00068}_{-0.00038}$ & 3 \\ 
\lbrack \ion{Fe}{7}\rbrack & 5158.890 & $3.44 \times 10^6$ & 99.00 & $^3$F$_3 - ^3$P$_1$ & $37.70 \pm 1.45$ & ${0.0025}^{+0.0020}_{-0.0012}$ & 4 \\ 
\lbrack \ion{Fe}{6}\rbrack & 5145.750 & $2.29 \times 10^7$ & 75.00 & $^4$F$_{7/2} - ^2$G$_{7/2}$ & $40.65 \pm 0.48$ & ${0.00077}^{+0.00041}_{-0.00028}$ & 7 \\ 
\lbrack \ion{Fe}{7}\rbrack & 4893.370 & $3.09 \times 10^6$ & 99.00 & $^3$F$_2 - ^3$P$_1$ & $38.94 \pm 2.21$ & ${0.0054}^{+0.0036}_{-0.0023}$ & 5 \\ 
\lbrack \ion{Fe}{5}\rbrack & 4180.600 & $1.86 \times 10^8$ & 54.80 & $^5$D$_1 - ^3$P2$_0$ & $39.45 \pm 2.15$ & ${0.00117}^{+0.00079}_{-0.00051}$ & 5 \\ 
\lbrack \ion{Fe}{5}\rbrack & 3891.280 & $1.61 \times 10^8$ & 54.80 & $^5$D$_4 - ^3$F2$_4$ & $40.62 \pm 1.06$ & ${0.00022}^{+0.00029}_{-0.00014}$ & 2 \\ 
\lbrack \ion{Fe}{5}\rbrack & 3839.270 & $1.00 \times 10^8$ & 54.80 & $^5$D$_3 - ^3$F2$_3$ & $41.20 \pm 0.60$ & ${0.00054}^{+0.00037}_{-0.00023}$ & 5 \\ 
\lbrack \ion{Fe}{7}\rbrack & 3758.920 & $4.02 \times 10^7$ & 99.00 & $^3$F$_4 - ^1$G$_4$ & $40.38 \pm 0.90$ & ${0.00097}^{+0.00048}_{-0.00033}$ & 8 \\ 
\lbrack \ion{Ne}{5}\rbrack & 3425.881 & $1.90 \times 10^7$ & 97.11 & $^3$P$_2 - ^1$D$_2$ & $41.21 \pm 0.29$ & ${0.0123}^{+0.0018}_{-0.0016}$ & 61 \\ 
\lbrack \ion{Ne}{5}\rbrack & 3345.821 & $1.14 \times 10^7$ & 97.11 & $^3$P$_1 - ^1$D$_2$ & $40.91 \pm 0.24$ & ${0.0085}^{+0.0018}_{-0.0015}$ & 31 \\ 
\enddata
\begin{tablenotes}
    \item[1] \textsuperscript{1}Wavelengths taken from: \url{https://physics.nist.gov/PhysRefData/ASD/lines_form.html}.
    \item[2] \textsuperscript{2}Ionization potential taken from: \url{https://physics.nist.gov/PhysRefData/ASD/ionEnergy.html}.
\end{tablenotes}
\label{tab:lines}
\end{deluxetable*}

%% file: main.bbl
\begin{thebibliography}{}
\providecommand\natexlab[1]{#1}
\providecommand\JournalTitle[1]{#1}

\bibitem[{{Abel} \& {Satyapal}(2008)}]{2008ApJ...678..686A}
{Abel}, N.~P., \& {Satyapal}, S. 2008,
  \href{http://dx.doi.org/10.1086/529013}{\JournalTitle{\apj}, 678, 686}

\bibitem[{{Aihara} {et~al.}(2011){Aihara}, {Allende Prieto}, {An}, {Anderson},
  {Aubourg}, {Balbinot}, {Beers}, {Berlind}, {Bickerton}, {Bizyaev}, {Blanton},
  {Bochanski}, {Bolton}, {Bovy}, {Brandt}, {Brinkmann}, {Brown}, {Brownstein},
  {Busca}, {Campbell}, {Carr}, {Chen}, {Chiappini}, {Comparat}, {Connolly},
  {Cortes}, {Croft}, {Cuesta}, {da Costa}, {Davenport}, {Dawson}, {Dhital},
  {Ealet}, {Ebelke}, {Edmondson}, {Eisenstein}, {Escoffier}, {Esposito},
  {Evans}, {Fan}, {Femen{\'\i}a Castell{\'a}}, {Font-Ribera}, {Frinchaboy},
  {Ge}, {Gillespie}, {Gilmore}, {Gonz{\'a}lez Hern{\'a}ndez}, {Gott}, {Gould},
  {Grebel}, {Gunn}, {Hamilton}, {Harding}, {Harris}, {Hawley}, {Hearty}, {Ho},
  {Hogg}, {Holtzman}, {Honscheid}, {Inada}, {Ivans}, {Jiang}, {Johnson},
  {Jordan}, {Jordan}, {Kazin}, {Kirkby}, {Klaene}, {Knapp}, {Kneib},
  {Kochanek}, {Koesterke}, {Kollmeier}, {Kron}, {Lampeitl}, {Lang}, {Le Goff},
  {Lee}, {Lin}, {Long}, {Loomis}, {Lucatello}, {Lundgren}, {Lupton}, {Ma},
  {MacDonald}, {Mahadevan}, {Maia}, {Makler}, {Malanushenko}, {Malanushenko},
  {Mandelbaum}, {Maraston}, {Margala}, {Masters}, {McBride}, {McGehee},
  {McGreer}, {M{\'e}nard}, {Miralda-Escud{\'e}}, {Morrison}, {Mullally},
  {Muna}, {Munn}, {Murayama}, {Myers}, {Naugle}, {Neto}, {Nguyen}, {Nichol},
  {O'Connell}, {Ogando}, {Olmstead}, {Oravetz}, {Padmanabhan},
  {Palanque-Delabrouille}, {Pan}, {Pandey}, {P{\^a}ris}, {Percival},
  {Petitjean}, {Pfaffenberger}, {Pforr}, {Phleps}, {Pichon}, {Pieri}, {Prada},
  {Price-Whelan}, {Raddick}, {Ramos}, {Reyl{\'e}}, {Rich}, {Richards}, {Rix},
  {Robin}, {Rocha-Pinto}, {Rockosi}, {Roe}, {Rollinde}, {Ross}, {Ross},
  {Rossetto}, {S{\'a}nchez}, {Sayres}, {Schlegel}, {Schlesinger}, {Schmidt},
  {Schneider}, {Sheldon}, {Shu}, {Simmerer}, {Simmons}, {Sivarani}, {Snedden},
  {Sobeck}, {Steinmetz}, {Strauss}, {Szalay}, {Tanaka}, {Thakar}, {Thomas},
  {Tinker}, {Tofflemire}, {Tojeiro}, {Tremonti}, {Vandenberg}, {Vargas
  Maga{\~n}a}, {Verde}, {Vogt}, {Wake}, {Wang}, {Weaver}, {Weinberg}, {White},
  {White}, {Yanny}, {Yasuda}, {Yeche}, \& {Zehavi}}]{2011ApJS..193...29A}
{Aihara}, H., {Allende Prieto}, C., {An}, D., {et~al.} 2011,
  \href{http://dx.doi.org/10.1088/0067-0049/193/2/29}{\JournalTitle{\apjs},
  193, 29}

\bibitem[{{Alexander} {et~al.}(2000){Alexander}, {Lutz}, {Sturm}, {Genzel},
  {Sternberg}, \& {Netzer}}]{2000ApJ...536..710A}
{Alexander}, T., {Lutz}, D., {Sturm}, E., {et~al.} 2000,
  \href{http://dx.doi.org/10.1086/308973}{\JournalTitle{\apj}, 536, 710}

\bibitem[{{Allen} {et~al.}(2008){Allen}, {Groves}, {Dopita}, {Sutherland}, \&
  {Kewley}}]{2008ApJS..178...20A}
{Allen}, M.~G., {Groves}, B.~A., {Dopita}, M.~A., {Sutherland}, R.~S., \&
  {Kewley}, L.~J. 2008,
  \href{http://dx.doi.org/10.1086/589652}{\JournalTitle{\apjs}, 178, 20}

\bibitem[{{Appenzeller} \& {Oestreicher}(1988)}]{1988AJ.....95...45A}
{Appenzeller}, I., \& {Oestreicher}, R. 1988,
  \href{http://dx.doi.org/10.1086/114611}{\JournalTitle{\aj}, 95, 45}

\bibitem[{{Astropy Collaboration} {et~al.}(2013){Astropy Collaboration},
  {Robitaille}, {Tollerud}, {Greenfield}, {Droettboom}, {Bray}, {Aldcroft},
  {Davis}, {Ginsburg}, {Price-Whelan}, {Kerzendorf}, {Conley}, {Crighton},
  {Barbary}, {Muna}, {Ferguson}, {Grollier}, {Parikh}, {Nair}, {Unther},
  {Deil}, {Woillez}, {Conseil}, {Kramer}, {Turner}, {Singer}, {Fox}, {Weaver},
  {Zabalza}, {Edwards}, {Azalee Bostroem}, {Burke}, {Casey}, {Crawford},
  {Dencheva}, {Ely}, {Jenness}, {Labrie}, {Lim}, {Pierfederici}, {Pontzen},
  {Ptak}, {Refsdal}, {Servillat}, \& {Streicher}}]{2013A&A...558A..33A}
{Astropy Collaboration}, {Robitaille}, T.~P., {Tollerud}, E.~J., {et~al.} 2013,
  \href{http://dx.doi.org/10.1051/0004-6361/201322068}{\JournalTitle{\aap},
  558, A33}

\bibitem[{{Baldwin} {et~al.}(1981){Baldwin}, {Phillips}, \&
  {Terlevich}}]{1981PASP...93....5B}
{Baldwin}, J.~A., {Phillips}, M.~M., \& {Terlevich}, R. 1981,
  \href{http://dx.doi.org/10.1086/130766}{\JournalTitle{\pasp}, 93, 5}

\bibitem[{{Benjamin} \& {Dinerstein}(1990)}]{1990AJ....100.1588B}
{Benjamin}, R.~A., \& {Dinerstein}, H.~L. 1990,
  \href{http://dx.doi.org/10.1086/115619}{\JournalTitle{\aj}, 100, 1588}

\bibitem[{{Bohn} {et~al.}(2021){Bohn}, {Canalizo}, {Veilleux}, \&
  {Liu}}]{2021ApJ...911...70B}
{Bohn}, T., {Canalizo}, G., {Veilleux}, S., \& {Liu}, W. 2021,
  \href{http://dx.doi.org/10.3847/1538-4357/abe70c}{\JournalTitle{\apj}, 911,
  70}

\bibitem[{{Boissay} {et~al.}(2016){Boissay}, {Ricci}, \&
  {Paltani}}]{2016A&A...588A..70B}
{Boissay}, R., {Ricci}, C., \& {Paltani}, S. 2016,
  \href{http://dx.doi.org/10.1051/0004-6361/201526982}{\JournalTitle{\aap},
  588, A70}

\bibitem[{{Brinchmann} {et~al.}(2004){Brinchmann}, {Charlot}, {White},
  {Tremonti}, {Kauffmann}, {Heckman}, \& {Brinkmann}}]{2004MNRAS.351.1151B}
{Brinchmann}, J., {Charlot}, S., {White}, S.~D.~M., {et~al.} 2004,
  \href{http://dx.doi.org/10.1111/j.1365-2966.2004.07881.x}{\JournalTitle{\mnras},
  351, 1151}

\bibitem[{{Campbell} {et~al.}(1986){Campbell}, {Terlevich}, \&
  {Melnick}}]{1986MNRAS.223..811C}
{Campbell}, A., {Terlevich}, R., \& {Melnick}, J. 1986,
  \href{http://dx.doi.org/10.1093/mnras/223.4.811}{\JournalTitle{\mnras}, 223,
  811}

\bibitem[{{Cann} {et~al.}(2019){Cann}, {Satyapal}, {Abel}, {Blecha},
  {Mushotzky}, {Reynolds}, \& {Secrest}}]{2019ApJ...870L...2C}
{Cann}, J.~M., {Satyapal}, S., {Abel}, N.~P., {et~al.} 2019,
  \href{http://dx.doi.org/10.3847/2041-8213/aaf88d}{\JournalTitle{\apjl}, 870,
  L2}

\bibitem[{{Cann} {et~al.}(2018){Cann}, {Satyapal}, {Abel}, {Ricci}, {Secrest},
  {Blecha}, \& {Gliozzi}}]{2018ApJ...861..142C}
---. 2018,
  \href{http://dx.doi.org/10.3847/1538-4357/aac64a}{\JournalTitle{\apj}, 861,
  142}

\bibitem[{{Cann} {et~al.}(2021){Cann}, {Satyapal}, {Rothberg}, {Canalizo},
  {Bohn}, {LaMassa}, {Matzko}, {Blecha}, {Secrest}, {Seth}, {B{\"o}ker},
  {Sexton}, {Kamal}, \& {Schmitt}}]{2021ApJ...912L...2C}
{Cann}, J.~M., {Satyapal}, S., {Rothberg}, B., {et~al.} 2021,
  \href{http://dx.doi.org/10.3847/2041-8213/abf56d}{\JournalTitle{\apjl}, 912,
  L2}

\bibitem[{{Cappellari}(2017)}]{2017MNRAS.466..798C}
{Cappellari}, M. 2017,
  \href{http://dx.doi.org/10.1093/mnras/stw3020}{\JournalTitle{\mnras}, 466,
  798}

\bibitem[{{Eisenstein} {et~al.}(2011){Eisenstein}, {Weinberg}, {Agol},
  {Aihara}, {Allende Prieto}, {Anderson}, {Arns}, {Aubourg}, {Bailey},
  {Balbinot}, {Barkhouser}, {Beers}, {Berlind}, {Bickerton}, {Bizyaev},
  {Blanton}, {Bochanski}, {Bolton}, {Bosman}, {Bovy}, {Brandt}, {Breslauer},
  {Brewington}, {Brinkmann}, {Brown}, {Brownstein}, {Burger}, {Busca},
  {Campbell}, {Cargile}, {Carithers}, {Carlberg}, {Carr}, {Chang}, {Chen},
  {Chiappini}, {Comparat}, {Connolly}, {Cortes}, {Croft}, {Cunha}, {da Costa},
  {Davenport}, {Dawson}, {De Lee}, {Porto de Mello}, {de Simoni}, {Dean},
  {Dhital}, {Ealet}, {Ebelke}, {Edmondson}, {Eiting}, {Escoffier}, {Esposito},
  {Evans}, {Fan}, {Femen{\'\i}a Castell{\'a}}, {Dutra Ferreira}, {Fitzgerald},
  {Fleming}, {Font-Ribera}, {Ford}, {Frinchaboy}, {Garc{\'\i}a P{\'e}rez},
  {Gaudi}, {Ge}, {Ghezzi}, {Gillespie}, {Gilmore}, {Girardi}, {Gott}, {Gould},
  {Grebel}, {Gunn}, {Hamilton}, {Harding}, {Harris}, {Hawley}, {Hearty},
  {Hennawi}, {Gonz{\'a}lez Hern{\'a}ndez}, {Ho}, {Hogg}, {Holtzman},
  {Honscheid}, {Inada}, {Ivans}, {Jiang}, {Jiang}, {Johnson}, {Jordan},
  {Jordan}, {Kauffmann}, {Kazin}, {Kirkby}, {Klaene}, {Knapp}, {Kneib},
  {Kochanek}, {Koesterke}, {Kollmeier}, {Kron}, {Lampeitl}, {Lang}, {Lawler},
  {Le Goff}, {Lee}, {Lee}, {Leisenring}, {Lin}, {Liu}, {Long}, {Loomis},
  {Lucatello}, {Lundgren}, {Lupton}, {Ma}, {Ma}, {MacDonald}, {Mack},
  {Mahadevan}, {Maia}, {Majewski}, {Makler}, {Malanushenko}, {Malanushenko},
  {Mandelbaum}, {Maraston}, {Margala}, {Maseman}, {Masters}, {McBride},
  {McDonald}, {McGreer}, {McMahon}, {Mena Requejo}, {M{\'e}nard},
  {Miralda-Escud{\'e}}, {Morrison}, {Mullally}, {Muna}, {Murayama}, {Myers},
  {Naugle}, {Neto}, {Nguyen}, {Nichol}, {Nidever}, {O'Connell}, {Ogando},
  {Olmstead}, {Oravetz}, {Padmanabhan}, {Paegert}, {Palanque-Delabrouille},
  {Pan}, {Pandey}, {Parejko}, {P{\^a}ris}, {Pellegrini}, {Pepper}, {Percival},
  {Petitjean}, {Pfaffenberger}, {Pforr}, {Phleps}, {Pichon}, {Pieri}, {Prada},
  {Price-Whelan}, {Raddick}, {Ramos}, {Reid}, {Reyle}, {Rich}, {Richards},
  {Rieke}, {Rieke}, {Rix}, {Robin}, {Rocha-Pinto}, {Rockosi}, {Roe},
  {Rollinde}, {Ross}, {Ross}, {Rossetto}, {S{\'a}nchez}, {Santiago}, {Sayres},
  {Schiavon}, {Schlegel}, {Schlesinger}, {Schmidt}, {Schneider}, {Sellgren},
  {Shelden}, {Sheldon}, {Shetrone}, {Shu}, {Silverman}, {Simmerer}, {Simmons},
  {Sivarani}, {Skrutskie}, {Slosar}, {Smee}, {Smith}, {Snedden}, {Stassun},
  {Steele}, {Steinmetz}, {Stockett}, {Stollberg}, {Strauss}, {Szalay},
  {Tanaka}, {Thakar}, {Thomas}, {Tinker}, {Tofflemire}, {Tojeiro}, {Tremonti},
  {Vargas Maga{\~n}a}, {Verde}, {Vogt}, {Wake}, {Wan}, {Wang}, {Weaver},
  {White}, {White}, {Wilson}, {Wisniewski}, {Wood-Vasey}, {Yanny}, {Yasuda},
  {Y{\`e}che}, {York}, {Young}, {Zasowski}, {Zehavi}, \&
  {Zhao}}]{2011AJ....142...72E}
{Eisenstein}, D.~J., {Weinberg}, D.~H., {Agol}, E., {et~al.} 2011,
  \href{http://dx.doi.org/10.1088/0004-6256/142/3/72}{\JournalTitle{\aj}, 142,
  72}

\bibitem[{{Erkens} {et~al.}(1997){Erkens}, {Appenzeller}, \&
  {Wagner}}]{1997A&A...323..707E}
{Erkens}, U., {Appenzeller}, I., \& {Wagner}, S. 1997, \JournalTitle{\aap},
  323, 707

\bibitem[{{Foreman-Mackey} {et~al.}(2013){Foreman-Mackey}, {Hogg}, {Lang}, \&
  {Goodman}}]{2013PASP..125..306F}
{Foreman-Mackey}, D., {Hogg}, D.~W., {Lang}, D., \& {Goodman}, J. 2013,
  \href{http://dx.doi.org/10.1086/670067}{\JournalTitle{\pasp}, 125, 306}

\bibitem[{Gelbord {et~al.}(2009)Gelbord, Mullaney, \&
  Ward}]{10.1111/j.1365-2966.2009.14961.x}
Gelbord, J.~M., Mullaney, J.~R., \& Ward, M.~J. 2009,
  \href{http://dx.doi.org/10.1111/j.1365-2966.2009.14961.x}{\JournalTitle{Monthly
  Notices of the Royal Astronomical Society}, 397, 172}

\bibitem[{{Goulding} \& {Alexander}(2009)}]{2009MNRAS.398.1165G}
{Goulding}, A.~D., \& {Alexander}, D.~M. 2009,
  \href{http://dx.doi.org/10.1111/j.1365-2966.2009.15194.x}{\JournalTitle{\mnras},
  398, 1165}

\bibitem[{{Grandi}(1978)}]{1978ApJ...221..501G}
{Grandi}, S.~A. 1978,
  \href{http://dx.doi.org/10.1086/156051}{\JournalTitle{\apj}, 221, 501}

\bibitem[{{Groves} {et~al.}(2006){Groves}, {Kewley}, {Kauffmann}, \&
  {Heckman}}]{2006NewAR..50..743G}
{Groves}, B., {Kewley}, L., {Kauffmann}, G., \& {Heckman}, T. 2006,
  \href{http://dx.doi.org/10.1016/j.newar.2006.06.081}{\JournalTitle{\nar}, 50,
  743}

\bibitem[{{Gunn} {et~al.}(2006){Gunn}, {Siegmund}, {Mannery}, {Owen}, {Hull},
  {Leger}, {Carey}, {Knapp}, {York}, {Boroski}, {Kent}, {Lupton}, {Rockosi},
  {Evans}, {Waddell}, {Anderson}, {Annis}, {Barentine}, {Bartoszek}, {Bastian},
  {Bracker}, {Brewington}, {Briegel}, {Brinkmann}, {Brown}, {Carr},
  {Czarapata}, {Drennan}, {Dombeck}, {Federwitz}, {Gillespie}, {Gonzales},
  {Hansen}, {Harvanek}, {Hayes}, {Jordan}, {Kinney}, {Klaene}, {Kleinman},
  {Kron}, {Kresinski}, {Lee}, {Limmongkol}, {Lindenmeyer}, {Long}, {Loomis},
  {McGehee}, {Mantsch}, {Neilsen}, {Neswold}, {Newman}, {Nitta}, {Peoples},
  {Pier}, {Prieto}, {Prosapio}, {Rivetta}, {Schneider}, {Snedden}, \&
  {Wang}}]{2006AJ....131.2332G}
{Gunn}, J.~E., {Siegmund}, W.~A., {Mannery}, E.~J., {et~al.} 2006,
  \href{http://dx.doi.org/10.1086/500975}{\JournalTitle{\aj}, 131, 2332}

\bibitem[{{Harris} {et~al.}(2020){Harris}, {Millman}, {van der Walt},
  {Gommers}, {Virtanen}, {Cournapeau}, {Wieser}, {Taylor}, {Berg}, {Smith},
  {Kern}, {Picus}, {Hoyer}, {van Kerkwijk}, {Brett}, {Haldane}, {del R{\'\i}o},
  {Wiebe}, {Peterson}, {G{\'e}rard-Marchant}, {Sheppard}, {Reddy}, {Weckesser},
  {Abbasi}, {Gohlke}, \& {Oliphant}}]{2020Natur.585..357H}
{Harris}, C.~R., {Millman}, K.~J., {van der Walt}, S.~J., {et~al.} 2020,
  \href{http://dx.doi.org/10.1038/s41586-020-2649-2}{\JournalTitle{\nat}, 585,
  357}

\bibitem[{{Izotov} \& {Thuan}(2009)}]{2009ApJ...707.1560I}
{Izotov}, Y.~I., \& {Thuan}, T.~X. 2009,
  \href{http://dx.doi.org/10.1088/0004-637X/707/2/1560}{\JournalTitle{\apj},
  707, 1560}

\bibitem[{{Izotov} {et~al.}(2021){Izotov}, {Thuan}, \&
  {Guseva}}]{2021MNRAS.508.2556I}
{Izotov}, Y.~I., {Thuan}, T.~X., \& {Guseva}, N.~G. 2021,
  \href{http://dx.doi.org/10.1093/mnras/stab2798}{\JournalTitle{\mnras}, 508,
  2556}

\bibitem[{{Izotov} {et~al.}(2012){Izotov}, {Thuan}, \&
  {Privon}}]{2012MNRAS.427.1229I}
{Izotov}, Y.~I., {Thuan}, T.~X., \& {Privon}, G. 2012,
  \href{http://dx.doi.org/10.1111/j.1365-2966.2012.22051.x}{\JournalTitle{\mnras},
  427, 1229}

\bibitem[{{Jarrett} {et~al.}(2011){Jarrett}, {Cohen}, {Masci}, {Wright},
  {Stern}, {Benford}, {Blain}, {Carey}, {Cutri}, {Eisenhardt}, {Lonsdale},
  {Mainzer}, {Marsh}, {Padgett}, {Petty}, {Ressler}, {Skrutskie}, {Stanford},
  {Surace}, {Tsai}, {Wheelock}, \& {Yan}}]{2011ApJ...735..112J}
{Jarrett}, T.~H., {Cohen}, M., {Masci}, F., {et~al.} 2011,
  \href{http://dx.doi.org/10.1088/0004-637X/735/2/112}{\JournalTitle{\apj},
  735, 112}

\bibitem[{{Kauffmann} {et~al.}(2003){Kauffmann}, {Heckman}, {Tremonti},
  {Brinchmann}, {Charlot}, {White}, {Ridgway}, {Brinkmann}, {Fukugita}, {Hall},
  {Ivezi{\'c}}, {Richards}, \& {Schneider}}]{2003MNRAS.346.1055K}
{Kauffmann}, G., {Heckman}, T.~M., {Tremonti}, C., {et~al.} 2003,
  \href{http://dx.doi.org/10.1111/j.1365-2966.2003.07154.x}{\JournalTitle{\mnras},
  346, 1055}

\bibitem[{{Kewley} {et~al.}(2001){Kewley}, {Dopita}, {Sutherland}, {Heisler},
  \& {Trevena}}]{2001ApJ...556..121K}
{Kewley}, L.~J., {Dopita}, M.~A., {Sutherland}, R.~S., {Heisler}, C.~A., \&
  {Trevena}, J. 2001,
  \href{http://dx.doi.org/10.1086/321545}{\JournalTitle{\apj}, 556, 121}

\bibitem[{{Krolik}(1999)}]{1999agnc.book.....K}
{Krolik}, J.~H. 1999, {Active galactic nuclei : from the central black hole to
  the galactic environment}

\bibitem[{{Lamperti} {et~al.}(2017){Lamperti}, {Koss}, {Trakhtenbrot},
  {Schawinski}, {Ricci}, {Oh}, {Landt}, {Riffel}, {Rodr{\'\i}guez-Ardila},
  {Gehrels}, {Harrison}, {Masetti}, {Mushotzky}, {Treister}, {Ueda}, \&
  {Veilleux}}]{2017MNRAS.467..540L}
{Lamperti}, I., {Koss}, M., {Trakhtenbrot}, B., {et~al.} 2017,
  \href{http://dx.doi.org/10.1093/mnras/stx055}{\JournalTitle{\mnras}, 467,
  540}

\bibitem[{{Lynden-Bell}(1969)}]{1969Natur.223..690L}
{Lynden-Bell}, D. 1969,
  \href{http://dx.doi.org/10.1038/223690a0}{\JournalTitle{\nat}, 223, 690}

\bibitem[{{McGaugh} {et~al.}(2017){McGaugh}, {Schombert}, \&
  {Lelli}}]{2017ApJ...851...22M}
{McGaugh}, S.~S., {Schombert}, J.~M., \& {Lelli}, F. 2017,
  \href{http://dx.doi.org/10.3847/1538-4357/aa9790}{\JournalTitle{\apj}, 851,
  22}

\bibitem[{{Molina} {et~al.}(2021{\natexlab{a}}){Molina}, {Reines}, {Greene},
  {Darling}, \& {Condon}}]{2021ApJ...910....5M}
{Molina}, M., {Reines}, A.~E., {Greene}, J.~E., {Darling}, J., \& {Condon},
  J.~J. 2021{\natexlab{a}},
  \href{http://dx.doi.org/10.3847/1538-4357/abe120}{\JournalTitle{\apj}, 910,
  5}

\bibitem[{{Molina} {et~al.}(2021{\natexlab{b}}){Molina}, {Reines}, {Latimer},
  {Baldassare}, \& {Salehirad}}]{2021ApJ...922..155M}
{Molina}, M., {Reines}, A.~E., {Latimer}, C.~J., {Baldassare}, V., \&
  {Salehirad}, S. 2021{\natexlab{b}},
  \href{http://dx.doi.org/10.3847/1538-4357/ac1ffa}{\JournalTitle{\apj}, 922,
  155}

\bibitem[{{Moorwood} {et~al.}(1996){Moorwood}, {Lutz}, {Oliva}, {Marconi},
  {Netzer}, {Genzel}, {Sturm}, \& {de Graauw}}]{1996A&A...315L.109M}
{Moorwood}, A.~F.~M., {Lutz}, D., {Oliva}, E., {et~al.} 1996,
  \JournalTitle{\aap}, 315, L109

\bibitem[{{M{\"u}ller-S{\'a}nchez} {et~al.}(2018){M{\"u}ller-S{\'a}nchez},
  {Hicks}, {Malkan}, {Davies}, {Yu}, {Shaver}, \&
  {Davis}}]{2018ApJ...858...48M}
{M{\"u}ller-S{\'a}nchez}, F., {Hicks}, E.~K.~S., {Malkan}, M., {et~al.} 2018,
  \href{http://dx.doi.org/10.3847/1538-4357/aab9ad}{\JournalTitle{\apj}, 858,
  48}

\bibitem[{{M{\"u}ller-S{\'a}nchez} {et~al.}(2011){M{\"u}ller-S{\'a}nchez},
  {Prieto}, {Hicks}, {Vives-Arias}, {Davies}, {Malkan}, {Tacconi}, \&
  {Genzel}}]{2011ApJ...739...69M}
{M{\"u}ller-S{\'a}nchez}, F., {Prieto}, M.~A., {Hicks}, E.~K.~S., {et~al.}
  2011,
  \href{http://dx.doi.org/10.1088/0004-637X/739/2/69}{\JournalTitle{\apj}, 739,
  69}

\bibitem[{{Mushotzky} \& {Ferland}(1984)}]{1984ApJ...278..558M}
{Mushotzky}, R., \& {Ferland}, G.~J. 1984,
  \href{http://dx.doi.org/10.1086/161822}{\JournalTitle{\apj}, 278, 558}

\bibitem[{{Nagao} {et~al.}(2000){Nagao}, {Taniguchi}, \&
  {Murayama}}]{2000AJ....119.2605N}
{Nagao}, T., {Taniguchi}, Y., \& {Murayama}, T. 2000,
  \href{http://dx.doi.org/10.1086/301411}{\JournalTitle{\aj}, 119, 2605}

\bibitem[{{Negus} {et~al.}(2021){Negus}, {Comerford}, {M{\"u}ller S{\'a}nchez},
  {Barrera-Ballesteros}, {Drory}, {Rembold}, \& {Riffel}}]{2021ApJ...920...62N}
{Negus}, J., {Comerford}, J.~M., {M{\"u}ller S{\'a}nchez}, F., {et~al.} 2021,
  \href{http://dx.doi.org/10.3847/1538-4357/ac1343}{\JournalTitle{\apj}, 920,
  62}

\bibitem[{{Netzer}(1985)}]{1985MNRAS.216...63N}
{Netzer}, H. 1985,
  \href{http://dx.doi.org/10.1093/mnras/216.1.63}{\JournalTitle{\mnras}, 216,
  63}

\bibitem[{{Nussbaumer} \& {Osterbrock}(1970)}]{1970ApJ...161..811N}
{Nussbaumer}, H., \& {Osterbrock}, D.~E. 1970,
  \href{http://dx.doi.org/10.1086/150585}{\JournalTitle{\apj}, 161, 811}

\bibitem[{{Oke} \& {Sargent}(1968)}]{1968ApJ...151..807O}
{Oke}, J.~B., \& {Sargent}, W. L.~W. 1968,
  \href{http://dx.doi.org/10.1086/149486}{\JournalTitle{\apj}, 151, 807}

\bibitem[{{Oliva} {et~al.}(1994){Oliva}, {Salvati}, {Moorwood}, \&
  {Marconi}}]{1994A&A...288..457O}
{Oliva}, E., {Salvati}, M., {Moorwood}, A.~F.~M., \& {Marconi}, A. 1994,
  \JournalTitle{\aap}, 288, 457

\bibitem[{{Penston} \& {Perez}(1984)}]{1984MNRAS.211P..33P}
{Penston}, M.~V., \& {Perez}, E. 1984,
  \href{http://dx.doi.org/10.1093/mnras/211.1.33P}{\JournalTitle{\mnras}, 211,
  33P}

\bibitem[{{Pogge} \& {Peterson}(1992)}]{1992AJ....103.1084P}
{Pogge}, R.~W., \& {Peterson}, B.~M. 1992,
  \href{http://dx.doi.org/10.1086/116127}{\JournalTitle{\aj}, 103, 1084}

\bibitem[{{Prieto} {et~al.}(2022){Prieto}, {Rodr{\'\i}guez-Ardila}, {Panda}, \&
  {Marinello}}]{2022MNRAS.510.1010P}
{Prieto}, A., {Rodr{\'\i}guez-Ardila}, A., {Panda}, S., \& {Marinello}, M.
  2022, \href{http://dx.doi.org/10.1093/mnras/stab3414}{\JournalTitle{\mnras},
  510, 1010}

\bibitem[{{Prieto} {et~al.}(2002){Prieto}, {P{\'e}rez Garc{\'\i}a}, \&
  {Rodr{\'\i}guez Espinosa}}]{2002MNRAS.329..309P}
{Prieto}, M.~A., {P{\'e}rez Garc{\'\i}a}, A.~M., \& {Rodr{\'\i}guez Espinosa},
  J.~M. 2002,
  \href{http://dx.doi.org/10.1046/j.1365-8711.2002.04985.x}{\JournalTitle{\mnras},
  329, 309}

\bibitem[{{Riffel} {et~al.}(2006){Riffel}, {Rodr{\'\i}guez-Ardila}, \&
  {Pastoriza}}]{2006A&A...457...61R}
{Riffel}, R., {Rodr{\'\i}guez-Ardila}, A., \& {Pastoriza}, M.~G. 2006,
  \href{http://dx.doi.org/10.1051/0004-6361:20065291}{\JournalTitle{\aap}, 457,
  61}

\bibitem[{{Rodr{\'\i}guez-Ardila} {et~al.}(2011){Rodr{\'\i}guez-Ardila},
  {Prieto}, {Portilla}, \& {Tejeiro}}]{2011ApJ...743..100R}
{Rodr{\'\i}guez-Ardila}, A., {Prieto}, M.~A., {Portilla}, J.~G., \& {Tejeiro},
  J.~M. 2011,
  \href{http://dx.doi.org/10.1088/0004-637X/743/2/100}{\JournalTitle{\apj},
  743, 100}

\bibitem[{{Rodr{\'\i}guez-Ardila} {et~al.}(2006){Rodr{\'\i}guez-Ardila},
  {Prieto}, {Viegas}, \& {Gruenwald}}]{2006ApJ...653.1098R}
{Rodr{\'\i}guez-Ardila}, A., {Prieto}, M.~A., {Viegas}, S., \& {Gruenwald}, R.
  2006, \href{http://dx.doi.org/10.1086/508864}{\JournalTitle{\apj}, 653, 1098}

\bibitem[{{Rodr{\'\i}guez-Ardila} {et~al.}(2002){Rodr{\'\i}guez-Ardila},
  {Viegas}, {Pastoriza}, \& {Prato}}]{2002ApJ...579..214R}
{Rodr{\'\i}guez-Ardila}, A., {Viegas}, S.~M., {Pastoriza}, M.~G., \& {Prato},
  L. 2002, \href{http://dx.doi.org/10.1086/342840}{\JournalTitle{\apj}, 579,
  214}

\bibitem[{{Satyapal} {et~al.}(2018){Satyapal}, {Abel}, \&
  {Secrest}}]{2018ApJ...858...38S}
{Satyapal}, S., {Abel}, N.~P., \& {Secrest}, N.~J. 2018,
  \href{http://dx.doi.org/10.3847/1538-4357/aab7f8}{\JournalTitle{\apj}, 858,
  38}

\bibitem[{{Satyapal} {et~al.}(2021){Satyapal}, {Kamal}, {Cann}, {Secrest}, \&
  {Abel}}]{2021ApJ...906...35S}
{Satyapal}, S., {Kamal}, L., {Cann}, J.~M., {Secrest}, N.~J., \& {Abel}, N.~P.
  2021, \href{http://dx.doi.org/10.3847/1538-4357/abbfaf}{\JournalTitle{\apj},
  906, 35}

\bibitem[{{Schaerer} \& {Stasi{\'n}ska}(1999)}]{1999A&A...345L..17S}
{Schaerer}, D., \& {Stasi{\'n}ska}, G. 1999, \JournalTitle{\aap}, 345, L17

\bibitem[{Sexton {et~al.}(2020)Sexton, Matzko, Darden, Canalizo, \&
  Gorjian}]{sexton_2020}
Sexton, R.~O., Matzko, W., Darden, N., Canalizo, G., \& Gorjian, V. 2020,
  \href{http://dx.doi.org/10.1093/mnras/staa3278}{\JournalTitle{Monthly Notices
  of the Royal Astronomical Society}, 500, 2871–2895}

\bibitem[{{Smee} {et~al.}(2013){Smee}, {Gunn}, {Uomoto}, {Roe}, {Schlegel},
  {Rockosi}, {Carr}, {Leger}, {Dawson}, {Olmstead}, {Brinkmann}, {Owen},
  {Barkhouser}, {Honscheid}, {Harding}, {Long}, {Lupton}, {Loomis}, {Anderson},
  {Annis}, {Bernardi}, {Bhardwaj}, {Bizyaev}, {Bolton}, {Brewington}, {Briggs},
  {Burles}, {Burns}, {Castander}, {Connolly}, {Davenport}, {Ebelke}, {Epps},
  {Feldman}, {Friedman}, {Frieman}, {Heckman}, {Hull}, {Knapp}, {Lawrence},
  {Loveday}, {Mannery}, {Malanushenko}, {Malanushenko}, {Merrelli}, {Muna},
  {Newman}, {Nichol}, {Oravetz}, {Pan}, {Pope}, {Ricketts}, {Shelden},
  {Sandford}, {Siegmund}, {Simmons}, {Smith}, {Snedden}, {Schneider},
  {SubbaRao}, {Tremonti}, {Waddell}, \& {York}}]{2013AJ....146...32S}
{Smee}, S.~A., {Gunn}, J.~E., {Uomoto}, A., {et~al.} 2013,
  \href{http://dx.doi.org/10.1088/0004-6256/146/2/32}{\JournalTitle{\aj}, 146,
  32}

\bibitem[{{Smith} {et~al.}(2009){Smith}, {Silverman}, {Chornock}, {Filippenko},
  {Wang}, {Li}, {Ganeshalingam}, {Foley}, {Rex}, \&
  {Steele}}]{2009ApJ...695.1334S}
{Smith}, N., {Silverman}, J.~M., {Chornock}, R., {et~al.} 2009,
  \href{http://dx.doi.org/10.1088/0004-637X/695/2/1334}{\JournalTitle{\apj},
  695, 1334}

\bibitem[{{Taylor}(2005)}]{2005ASPC..347...29T}
{Taylor}, M.~B. 2005, in Astronomical Society of the Pacific Conference Series,
  Vol. 347, Astronomical Data Analysis Software and Systems XIV, ed.
  P.~{Shopbell}, M.~{Britton}, \& R.~{Ebert}, 29

\bibitem[{{Trump} {et~al.}(2015){Trump}, {Sun}, {Zeimann}, {Luck}, {Bridge},
  {Grier}, {Hagen}, {Juneau}, {Montero-Dorta}, {Rosario}, {Brandt},
  {Ciardullo}, \& {Schneider}}]{2015ApJ...811...26T}
{Trump}, J.~R., {Sun}, M., {Zeimann}, G.~R., {et~al.} 2015,
  \href{http://dx.doi.org/10.1088/0004-637X/811/1/26}{\JournalTitle{\apj}, 811,
  26}

\bibitem[{{Virtanen} {et~al.}(2020){Virtanen}, {Gommers}, {Oliphant},
  {Haberland}, {Reddy}, {Cournapeau}, {Burovski}, {Peterson}, {Weckesser},
  {Bright}, {van der Walt}, {Brett}, {Wilson}, {Millman}, {Mayorov}, {Nelson},
  {Jones}, {Kern}, {Larson}, {Carey}, {Polat}, {Feng}, {Moore}, {VanderPlas},
  {Laxalde}, {Perktold}, {Cimrman}, {Henriksen}, {Quintero}, {Harris},
  {Archibald}, {Ribeiro}, {Pedregosa}, {van Mulbregt}, \& {SciPy 1. 0
  Contributors}}]{2020NatMe..17..261V}
{Virtanen}, P., {Gommers}, R., {Oliphant}, T.~E., {et~al.} 2020,
  \href{http://dx.doi.org/10.1038/s41592-019-0686-2}{\JournalTitle{Nature
  Methods}, 17, 261}

\bibitem[{{Wang} {et~al.}(2012){Wang}, {Zhou}, {Komossa}, {Wang}, {Yuan}, \&
  {Yang}}]{2012ApJ...749..115W}
{Wang}, T.-G., {Zhou}, H.-Y., {Komossa}, S., {et~al.} 2012,
  \href{http://dx.doi.org/10.1088/0004-637X/749/2/115}{\JournalTitle{\apj},
  749, 115}

\bibitem[{{Wright} {et~al.}(2010){Wright}, {Eisenhardt}, {Mainzer}, {Ressler},
  {Cutri}, {Jarrett}, {Kirkpatrick}, {Padgett}, {McMillan}, {Skrutskie},
  {Stanford}, {Cohen}, {Walker}, {Mather}, {Leisawitz}, {Gautier}, {McLean},
  {Benford}, {Lonsdale}, {Blain}, {Mendez}, {Irace}, {Duval}, {Liu}, {Royer},
  {Heinrichsen}, {Howard}, {Shannon}, {Kendall}, {Walsh}, {Larsen}, {Cardon},
  {Schick}, {Schwalm}, {Abid}, {Fabinsky}, {Naes}, \&
  {Tsai}}]{2010AJ....140.1868W}
{Wright}, E.~L., {Eisenhardt}, P. R.~M., {Mainzer}, A.~K., {et~al.} 2010,
  \href{http://dx.doi.org/10.1088/0004-6256/140/6/1868}{\JournalTitle{\aj},
  140, 1868}

\bibitem[{{Zhang} {et~al.}(2012){Zhang}, {Fang}, {Chau}, {Hsia}, {Liu}, {Kwok},
  \& {Koning}}]{2012ApJ...754...28Z}
{Zhang}, Y., {Fang}, X., {Chau}, W., {et~al.} 2012,
  \href{http://dx.doi.org/10.1088/0004-637X/754/1/28}{\JournalTitle{\apj}, 754,
  28}

\end{thebibliography}
